\documentclass{aastex62}

\shorttitle{FINMHD}
\shortauthors{Baty et al.}

\usepackage{bm}%
\usepackage{graphicx}%
\usepackage{amsmath}

\def\ltsima{$\; \buildrel < \over \sim \;$}
\def\gtsima{$\; \buildrel > \over \sim \;$}
\def\simlt{\lower.5ex\hbox{\ltsima}}
\def\simgt{\lower.5ex\hbox{\gtsima}}

\begin{document}

\title{FINMHD: an adaptive finite element code for magnetic reconnection and plasmoid chains formation in Magnetohydrodynamics}%

\correspondingauthor{Hubert BATY }
\email{hubert.baty@astro.unistra.fr \\ \\
Submitted to APJS}

\author{ Hubert BATY}
\affiliation{Observatoire Astronomique de Strasbourg, Universit\'e de Strasbourg, CNRS, UMR 7550, 11 rue de l'Universit\'e, F-67000 Strasbourg, France}



\begin{abstract}
Solving the problem of fast eruptive events in magnetically dominated astrophysical plasmas requires the
use of particularly well adapted numerical tools. Indeed, the central mechanism based on magnetic reconnection is determined
by a complex behavior with quasi-singular forming current layers enriched by their associated 
small scale magnetic islands called plasmoids. A new code for the solution of two dimensional dissipative magnetohydrodynamics (MHD)
equations in cartesian geometry specifically developed to this end is thus presented.
A current-vorticity formulation representative of an incompressible model is chosen in order to follow the formation of the
current sheets and the ensuing magnetic reconnection process. A finite element discretization using triangles with quadratic
basis functions on an unstructured grid is employed, and implemented via a highly adaptive characteristic-Galerkin scheme.
The adaptivity of the code is illustrated on simplified test equations and finally for magnetic reconnection associated to
the non linear development of the tilt instability between two repelling current channels. Varying the Lundquist
number $S$,  has allowed to study the transition between the steady-state Sweet-Parker reconnection regime (for $S \simlt 10^4$) and plasmoids dominated
reconnection one (for $S \simgt 10^5$). The implications for the understanding of the mechanism explaining
the fast  conversion of free magnetic energy in astrophysical environments such as in solar corona are briefly discussed.
\end{abstract}


\keywords{magnetic reconnection --- magnetohydrodynamics ---  plasmas --- stars: coronae --- Sun: flares}


\section{Introduction} \label{sec:intro}

Magnetic reconnection is a fundamental process in laboratory and space plasmas, which allows the conversion of magnetic energy into
bulk flow and heating. The understanding of magnetic reconnection is a major goal of theoretical plasma physics in order to explain
explosive events like solar/stellar flares, coronal mass ejections, and gamma-ray flares in pulsar winds. For collisional plasmas, the
magnetohydrodynamic (MHD) model is commonly adopted as an excellent framework \citep{pri00}.
The MHD approximation is a single fluid description of plasma dynamics, where the electromagnetic properties are taken into account
via coupling terms obeying Maxwell's equations. Typically, electrical currents can be generated by via fluid motions producing
Lorentz forces on the fluid. The plasma resistivity (inverse of the electrical conductivity) representing the magnetic field dissipation, i.e. $\eta$,
is very small in plasmas of interest, but it cannot be neglected as it precisely drives the reconnection process. 
The classical model of reconnection is based on Sweet-Parker theory in the two-dimensional (2D) resistive MHD framework, in which a steady-state current
sheet structure with a small central diffusion layer controls the reconnection between two regions of oppositely directed magnetic
fields \citep{swe58,1957JGR...62...509}. However, the Sweet-Parker (hereafter SP) model gives a reconnection rate scaling as $\eta^{1/2}$, that is
too small by a few orders of magnitude to explain the fast time scales involved in the previously mentioned eruptive events.

After many years of research on the existence of viable alternate models, it has become recently clear that the problem can be solved by 
obtaining a new regime of accelerated reconnection when the resistivity is low enough \citep{2007PhPl...14j0703L}.
More explicitely, this is the case in plasmas having local Lundquist numbers $S$ higher than a critical value $S_c$ that is order $10^4$, where
$S = l V_A/\eta$ is defined by taking $l$ as the characteristic length scale of the current sheet and $V_A$ the characteristic
Alfv\'en speed based on the local magnetic field just upstream off the layer. As shown by many numerical MHD simulations,
this regime is characterized by a current layer broken into a chain of smaller magnetic islands called plasmoids, constantly
forming, moving, eventually coalescing, and finally being ejected through the outflow boundary. At a given time, the system appears
as an aligned layer structure of plasmoids of different sizes, and can be regarded as a statistical steady state with an average
reconnection rate that is nearly (or exactly) independent of the resistivity \citep{2009PhPl...16k2102B, 2009PhRvL.103j5004S, 2017ApJ...849...75H}.

There are however some fundamental issues about the growth of the plasmoids and the ensuing reconnection rate 
that remain unsolved, reflecting our inadequate understanding of the mechanism.
For example, a model based on the modal stability analysis of Sweet-Parker forming current sheets, proposed by \citet{2017ApJ...850..142C},
has led to scaling laws of the plasmoid instability growth that are not simple power laws with the resistivity (or Lundquist number).
Moreover, the fully plasmoid dominated regime is thus predicted to be obtained when $S$ is higher than another critical Lundquist $S_T$,
called the transitional Lundquist number, that is substantially  higher than $S_c$. For example, a value of $S_T \simeq 10^6 - 10^7$
is estimated for typical solar corona parameters. This transition corresponds to the inequality $\tau_p \simlt \tau$, where
$\tau_p$ and $\tau$ represent the characteristic time scales of the plasmoids growth and of the SP current sheet formation (that is for example
an exponentially shrinking process due to some MHD instability) respectively.
The latter results are in apparent contradiction with another stability study \citep{2014ApJ...780L..19P}, where the linear growth rate of the plasmoids
scales as $S^{1/4}$, and where the divergence for an infinite Lundquist number is regularized by considering
current sheets having asymptotic aspect ratios, $l/a \simeq S^{1/3}$ ($a$ being the characteristic width scale of the current layer), that are smaller than Sweet-Parker ones (i.e. equal to $S^{1/2}$).
This second model also predicts constant $\tau_p  \simeq l/V_A$ independent of $S$ in the limit of infinite $S$.

Most of the previous numerical studies have used general MHD codes, though not always
specifically designed for investigating the magnetic reconnection process. This is for example the case of finite volume based codes with shock-capturing
methods using different Riemann-type solvers in order to handle discontinuities and shocks, as for example in AMRVAC (see  \citet{2014ApJS..214....4P}
for code description, and \citet{2017ApJ...837...74B} for an example of application).
On the other hand, despite the fact that they are probably the best adapted to such problems, one finds only a few attempts in
the literature of using codes based on finite element methods. The exceptions concern mainly the related topic of the formation
of singularities in ideal MHD \citep{1998JCoPh.147..318S, 2007JCoPh.225..363L}, but not the reconnection process itself.
Conventional codes lack some convergence properties in the low resistivity regime, which is characterized by a complicated time dependent
bursty dynamics \citep{2013PhPl...20i2109K}. We thus develop a new code, FINMHD, using a finite element discretization in order to specifically tackle this
problem.

Most of the previous numerical studies on the subject also consider the same initial setup, namely the coalescence instability between two attracting current channels
(see \citet{2017ApJ...849...75H} for example).
This is a convenient configuration for numerical treatment, that has however the following drawbacks. Indeed, an initial
current layer of arbitrary width is assumed between the two channels. Second, this is important to use other configurations in order to address the generality
of the mechanism, with cases involving different current sheets formation and associated time scales.
Consequently, we have chosen to consider a different setup, that is the tilt instability between two repelling currents \citep{1990PhFlB...2..488R}.
Curiously, this latter configuration has been used to test ideal MHD codes \citep{2007JCoPh.225..363L}, or more recently to generate complex
magnetic structures to study particles acceleration \citep{2014ApJ...795...77K, 2017MNRAS.467.3279R}, but not to study magnetic reconnection.

The outline of the paper is as follows. In Section 2, we present our simplified model of 2D incompressible MHD equations written in current-vorticity variables,
usually called reduced MHD in literature. We also define two simplified model equations (advection diffusion and Burgers equations),
which are designed for testing the essential features of our numerical method. This is followed by the description of our characteristic-Galerkin
method applied to the simplified equations in Section 3, and the illustration of the efficiency of our adaptive scheme to capture the formation and evolution
of nearly singular structures on three examples in Section 4. In Section 5, we present the FINMHD code and its application on the evolution of the tilt instability. We also 
investigate different reconnection regimes by varying the resistivity parameter.
Finally, future perspectives using FINMHD, implications for the understanding of the reconnection mechanism in eruptive solar events, and conclusions
are reported in Section 6.

\section{Model equations}

\subsection{Reduced MHD Model}
We first consider the 2D incompressible set of dissipative (viscous and resistive) MHD equations written in flux-vorticity ($\psi-\omega$) scalar variables
as follows,
\begin{equation}
        \frac{\partial \omega}{\partial t} + (\bm{V}\cdot\bm{\nabla})\omega =  (\bm{B}\cdot\bm{\nabla}) J + \nu \bm{\nabla}^2 \omega ,
\end{equation}
\begin{equation}
    \frac{\partial \psi}{\partial t} + (\bm{V}\cdot\bm{\nabla})\psi = \eta \bm{\nabla}^2 \psi ,
\end{equation}
\begin{equation}
    \bm{\nabla}^2\phi = - \omega ,
\end{equation}
\begin{equation}
  \bm{\nabla}^2\psi = - J ,
    \end{equation}
where we have introduced the two stream functions, $\phi (x, y)$ and $\psi (x, y)$, defined as $\bm{V} = {\nabla} \phi \wedge \bm{e_z}$ and $\bm{B} = {\nabla} \psi \wedge \bm{e_z}$ ($\bm{e_z}$
being the unit vector perpendicular to the $xOy$ simulation plane).
Note that $J$ and $\omega$ are the $z$ components of the current density and vorticity vectors, as $\bm{J} = \nabla \wedge \bm{B}$ and $\bm{ \omega} = \nabla \wedge \bm{V}$ respectively (with units using $\mu_0 = 1$). Note that we consider the resistive diffusion via the $\eta \bm{\nabla}^2 \psi $ term ($\eta$ being assumed uniform for simplicity), and also a viscous term
$\nu \bm{\nabla}^2 \phi$ in a similar way (with $\nu$ being the viscosity parameter).
The above definitions results from the choice $\psi \equiv A_z$, where $A_z$ is the $z$ component of the potentiel vector $\bm{A}$ (as $\bm{B} = \nabla \wedge \bm{A}$). This choice
is the one used in \citet{2008ApJS..177..613N} or in  \citet{2016MNRAS.459..624B},
and different from the one used by \citet{2007JCoPh.225..363L} where the choice $\psi \equiv - A_z$ is done.
In the latter case, the two Poisson equations (i.e. Equations 3-4) involve an opposite sign in the right hand sides. Note that the thermal pressure gradient is naturally absent from our set of equations. 
The main advantage of the above formulation over a standard one using the velocity and magnetic field vectors ($\bm{V}, \bm{B}$)  as
the main variables, is the divergence-free property naturally ensured for these two vectors. However, except when spectral/pseudo-spectral methods
are employed, the space discretization of the flux-vorticity formulation is known to generate cell interfaces numerical instabilities due to the evaluation of the
third order spatial derivative term $(\bm{B}\cdot\bm{\nabla})J $, as $J$ is itself deduced from $\bm{\nabla}^2\psi = - J$ \citep{phi07}.

Fortunately, another formulation using current-vorticity ($J-\omega$) variables cures the above difficulty. The latter is obtained by
taking the Laplacien of the evolution equation for $\psi$, leading to a new equation for $J$ (now the maximum
spatial derivative is second order),
\begin{equation}  
      \frac{\partial \omega}{\partial t} + (\bm{V}\cdot\bm{\nabla})\omega = (\bm{B}\cdot\bm{\nabla})J + \nu \bm{\nabla}^2 \omega ,
\end{equation}
\begin{equation}      
        \frac{\partial J }{\partial t} + (\bm{V}\cdot\bm{\nabla})J =  (\bm{B}\cdot\bm{\nabla})\omega + \eta \bm{\nabla}^2 J +  g(\phi,\psi) ,
\end{equation}
\begin{equation}                 
     \bm{\nabla}^2\phi = - \omega ,
 \end{equation}
\begin{equation}                        
     \bm{\nabla}^2\psi = - J ,  
\end{equation}
with $g(\phi,\psi)=2 \left[\frac{\partial^2 \phi}{\partial x\partial y}\left(\frac{\partial^2 \psi}{\partial x^2} - \frac{\partial^2 \psi}{\partial y^2}\right) - \frac{\partial^2 \psi}{\partial x\partial y}\left(\frac{\partial^2 \phi}{\partial x^2} - \frac{\partial^2 \phi}{\partial y^2}\right)\right]$. 
Note that this second formulation has another advantage over the first one, as the two evolution equations are more symmetric, facilitating thus the
numerical matrices calculus (see below) for linear solvers.

\subsection{Simplified 2D advection-diffusion and viscous Burgers equations}
For the sake of introducing the basic concepts of our numerical method and testing its essential properties, we also consider two
simplified 2D time dependent equations. 

First, we consider an advection-diffusion equation for the variable $u(x,y,t)$,
\begin{equation}
        \frac{\partial u}{\partial t} + (\bm{V}\cdot\bm{\nabla})u =  \nu \bm{\nabla}^2 u + f(u) ,
\end{equation}
where $\bm{V} = (V_x, V_y)$ is a given velocity, $\nu$ is the viscosity parameter, and $f(u)$ is a general
source term that can be non linear.

Second, we consider the following viscous Burgers equation for the variable $u(x,y,t)$,
\begin{equation}
        \frac{\partial u}{\partial t} + u \frac{\partial u}{\partial x} + V_y  \frac{\partial u}{\partial y} =  \nu \bm{\nabla}^2 u + f(u) ,
\end{equation}
where $V_y$ is a given velocity along the $y$ direction, and $\nu$ is the viscosity parameter. Note that the quasi-linear term
involving shocks ($ u \frac{\partial u}{\partial x}$) is one dimensional. This second equation can be also written in a similar way
as the previous equation (Equation 9) with the velocity vector, $\bm{V} = (u, V_y)$. 

\begin{figure}
\centering
 \includegraphics[scale=0.46]{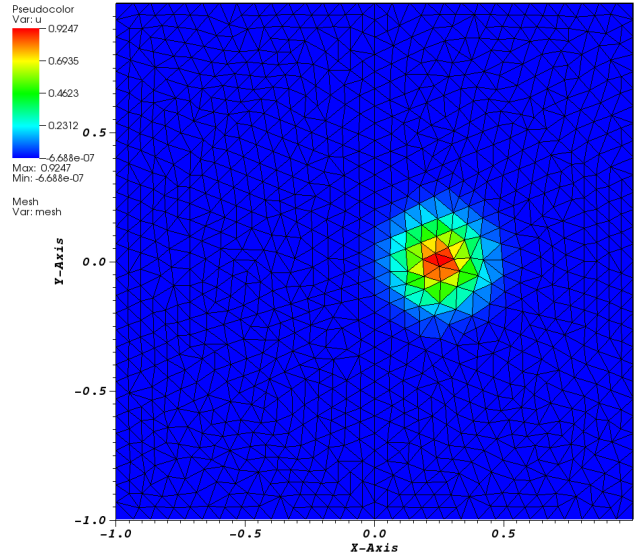}
  \caption{Approximate initial bell solution (color contour map) on a uniform crude mesh 
  with $2116$ triangles and $P_1$ elements.}
\end{figure}

\begin{figure}
\centering
 \includegraphics[scale=0.33]{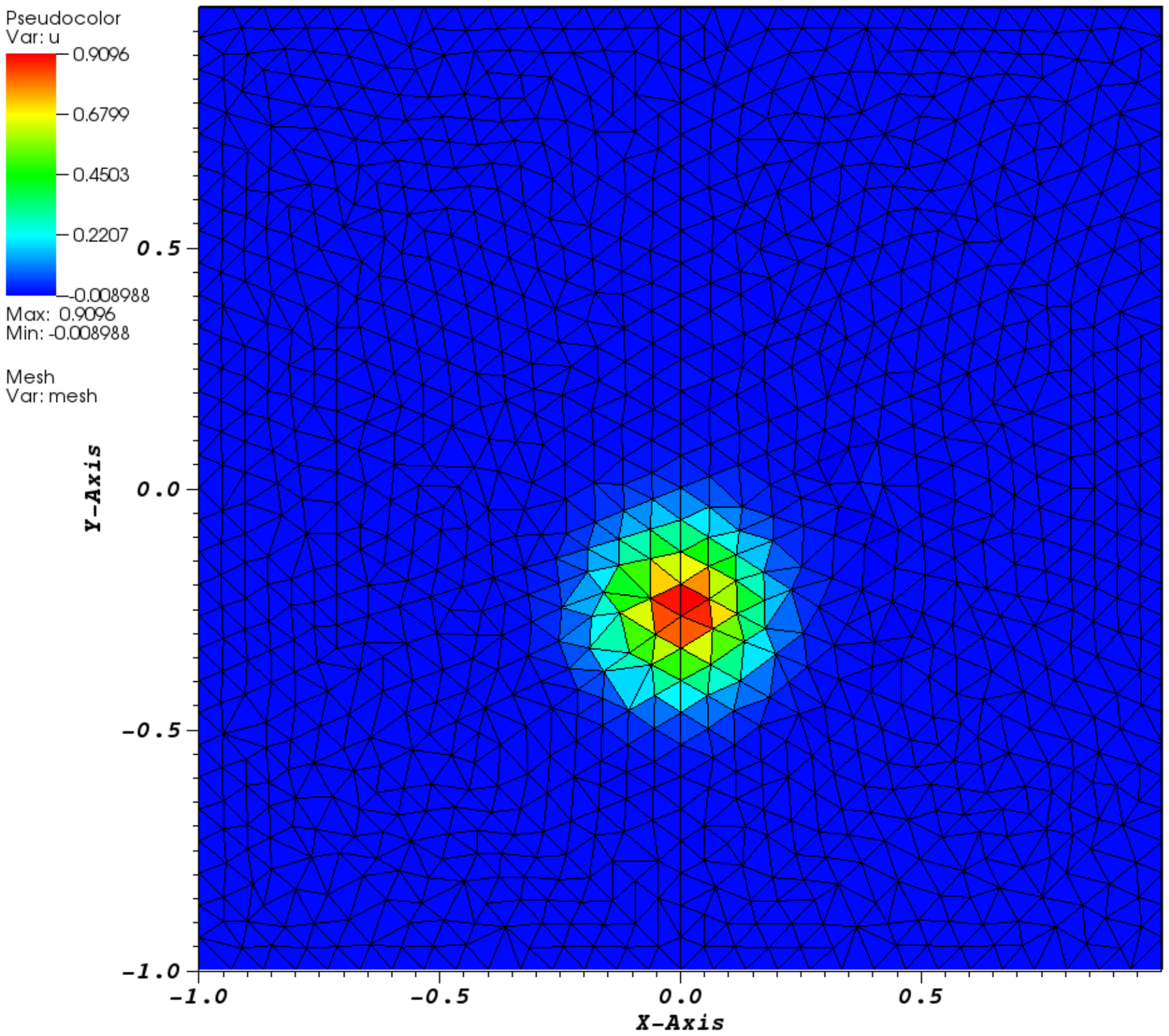}
 \includegraphics[scale=0.33]{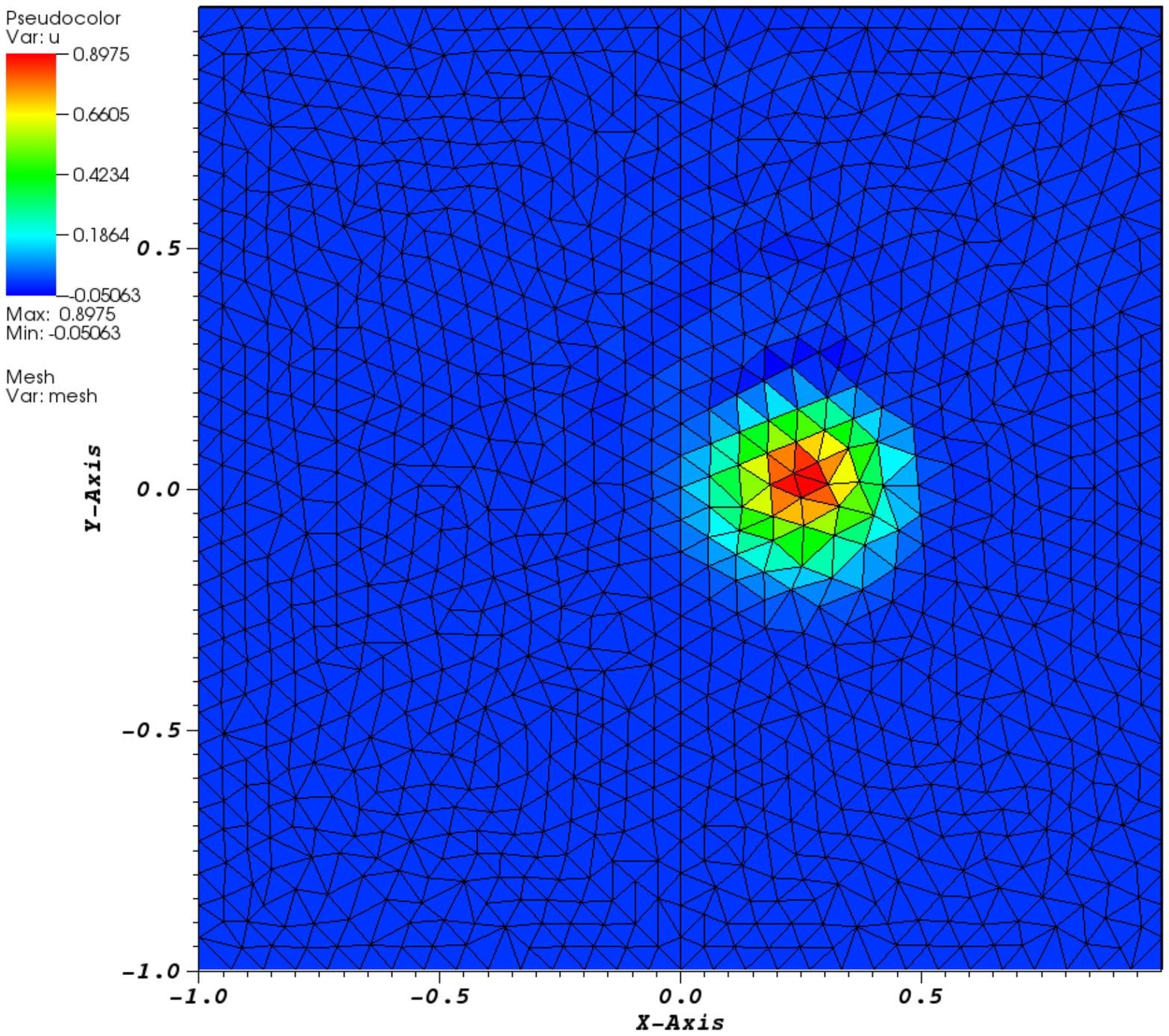}
  \caption{Approximate bell solution obtained on a crude uniform mesh with $2116$ triangles and $P_1$ elements
  (see previous figure) for a final time corresponding to a quarter of a turn (left panel), and to a full turn (right panel). }
\end{figure}

\begin{figure}
\centering
 \includegraphics[scale=0.33]{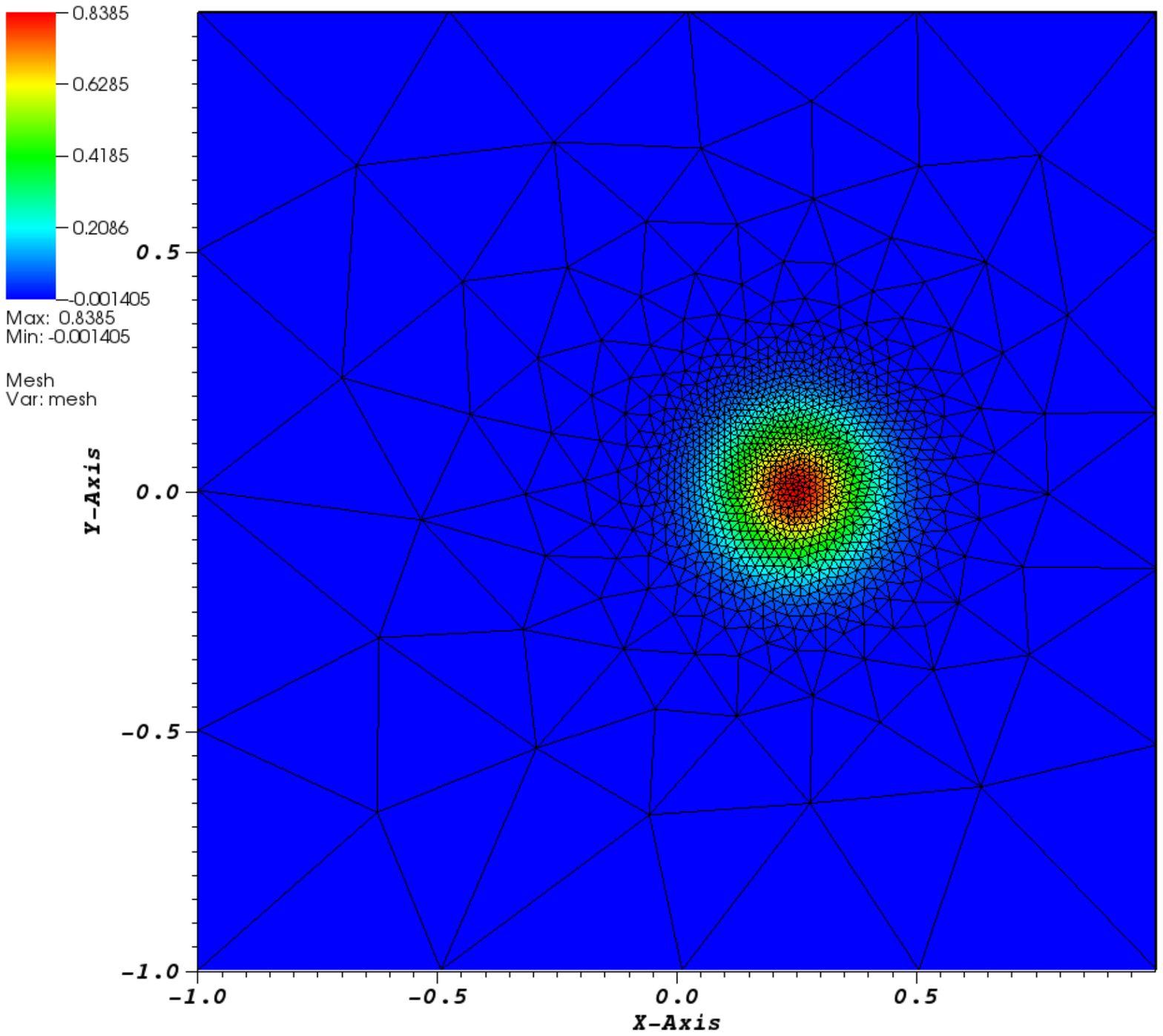}
 \includegraphics[scale=0.33]{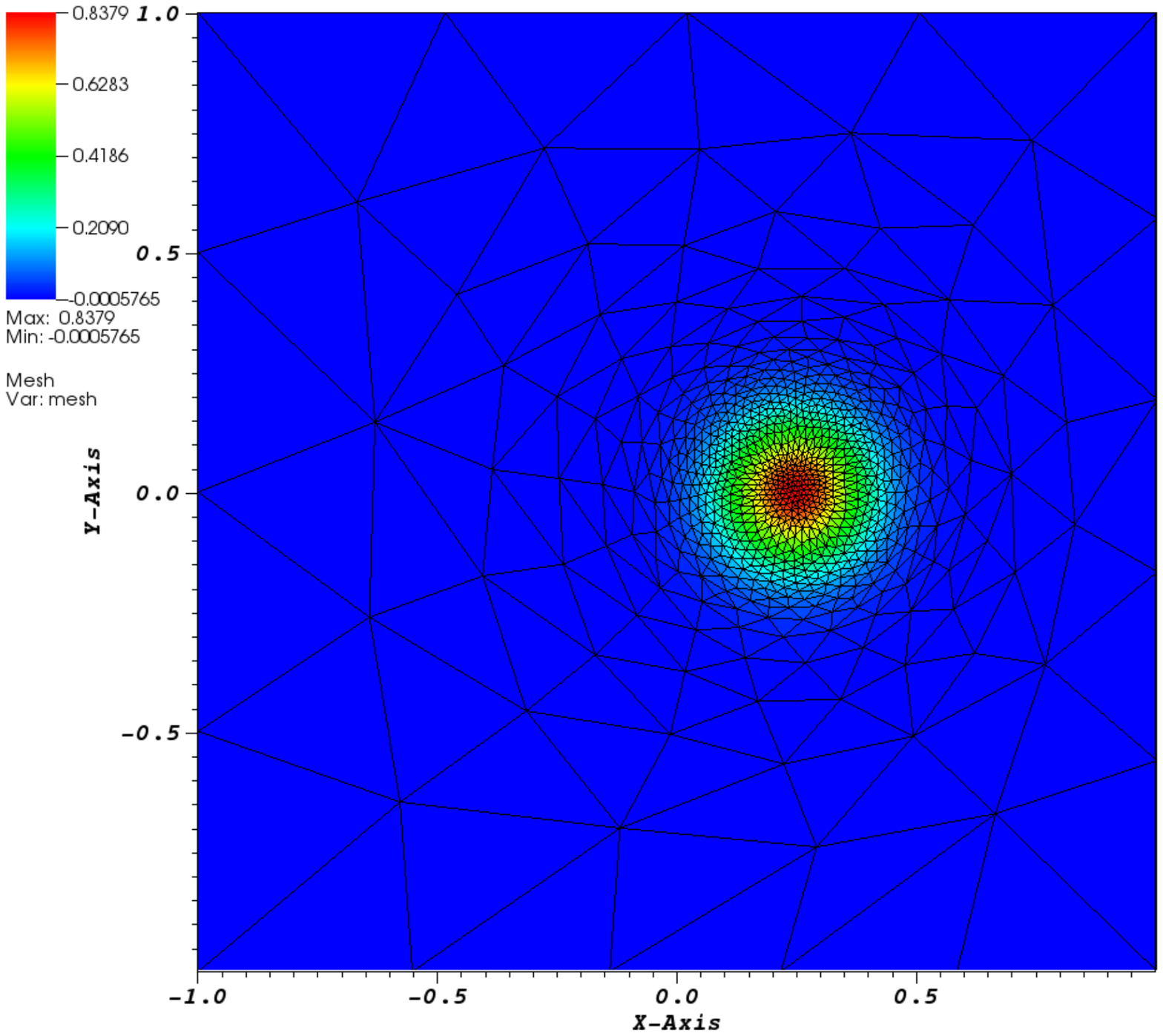}
  \caption{Approximate bell solution obtained for a full turn, using an isotropic adaptive scheme with $2200$ (in fact varying between $3200$ and $2200$)
  $P_1$ elements (left panel), and a non-isotropic adaptive variant (with a number of elements varying between $1900$ and $1400$)  (right panel).
  In both cases, the maximum triangle edge size $h_{max}$ is fixed to a value close to $0.5$.}
\end{figure}

\begin{figure}
\centering
 \includegraphics[scale=0.5]{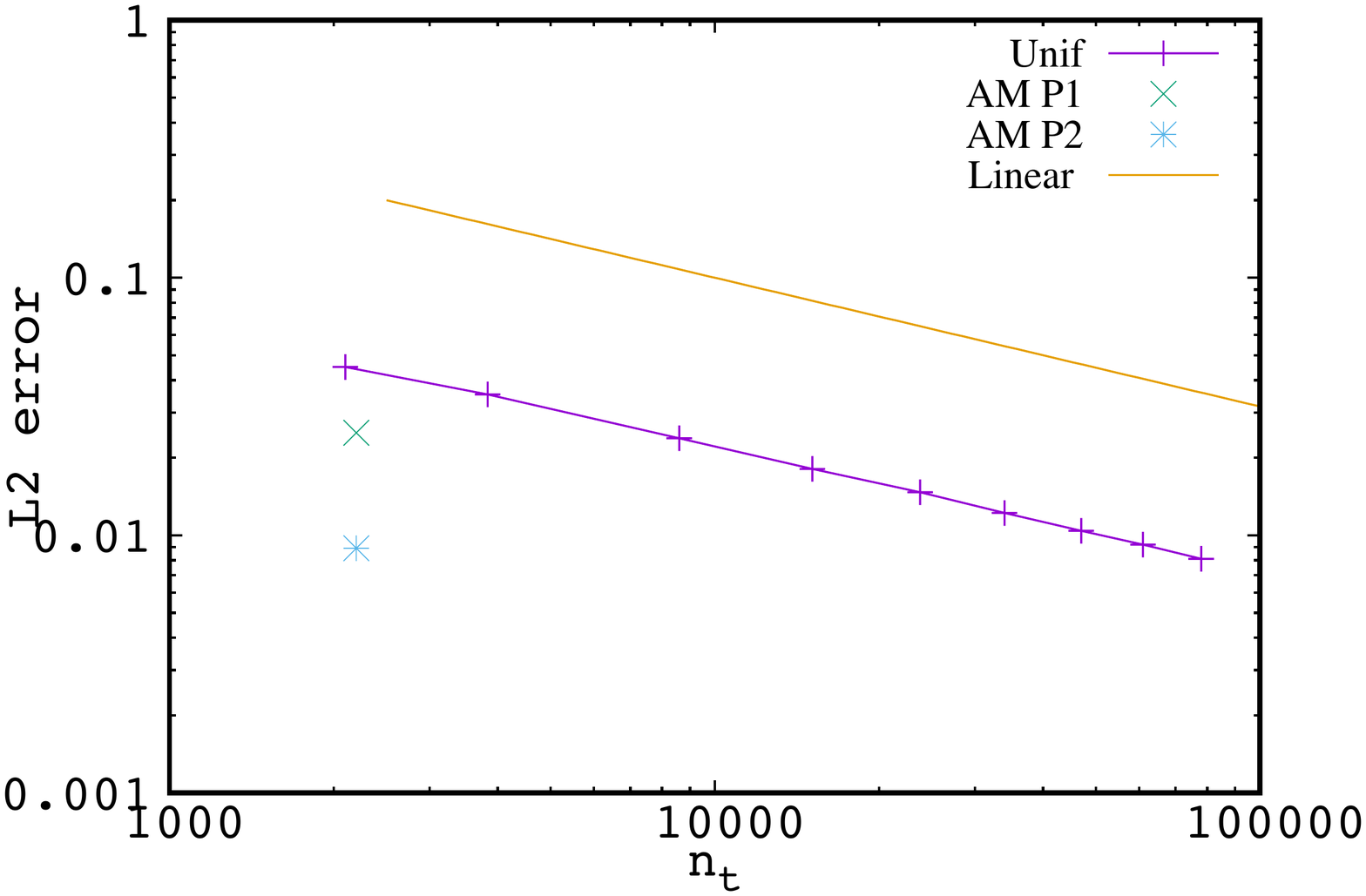}
  \caption{$L^2$ error norm obtained for the Gaussian bell rotation (full turn) using uniform meshes as a function of the number of triangles $n_t$ (Unif). A linear
  variation with $n_t^{0.5}$ is plotted. The $L^2$ errors obtained using the adaptive scheme with isotropic $P_1$ elements (AM P1)
  and isotropic $P_2$ elements (AM P2) are also reported. Note that the number of triangles and the CPU time is slightly lower for $P_1$ versus $P_2$.
  Using a direct solver is also slightly faster.}
\end{figure}

\section{The adaptive characteristic-Galerkin finite element method for the simplified equations}

\subsection{The standard Galerkin method}
The standard Galerkin method is obtained by multiplying the relevant equation to be solved by a test function $v_h (x, y)$
defined on a partition $\tau_h$ of the domain $\tau$ (with a mesh of non-overlapping elements of size $h$ that can be triangles for example),
and integrating over the whole domain. This leads to the following Galerkin form for the previous test equation,
\begin{equation}
 \left(\frac{\partial u_h}{\partial t} + (\bm{V}\cdot\bm{\nabla}) u_h ~,~ v_h\right) + \nu\left(\bm{\nabla} u_h ~,~\bm{\nabla} v_h \right) 
        -   \left (f(u_h) ~,~  v_h \right)= 0 ,
 \end{equation}
 where $u_h$ is the weak form solution on $\tau_h$ approximating $u$ on $\tau$, and where we use the definition
$\left(u_h~,~v_h \right) \equiv  \int_{\tau_h} u_h v_h \,d\tau$ evaluated as $\sum\limits_{k=1}^{N}  \int_{T_k} u_h v_h d\tau$
($N$ being the number of elements $T_k$ covering the partition $\tau_h$).
Note that the Laplacian term is now expressed as a bilinear form, as the result of using
mathematical Green's theorem and assuming homogeneous Dirichlet boundary conditions for simplicity. Additional integral
term  (in Equation 11) along the domain boundary must be also added taking into account eventual Neumann conditions, and additional Dirichlet conditions
can be also added separately.

The second step in Galerkin method concerns the choice of the elements (finite element functions) in order to cover the domain. In this work, the planar domain
is triangulated using Lagrange basis with polynomial functions of order 1 (usually called $P_1$ elements), or of order 2 ($P_2$ elements).
The variable $u_h$ can be thus expanded as, $u_h = \sum\limits_{i=1}^{M}  u_i  \lambda_i (x,y)$, where $M$ is the dimension
of the approximating space and $\lambda_i (x,y)$ are the usual nodal basis functions. Substituting the above expression into the equation, and replacing the
test function by each function of the basis, results in an algebraic system for the unknown coefficient $u_i$.
More precisely, for the vector $\overline{u}$ containing the coefficients, we have,
\begin{equation}
M. \frac{\partial\overline{u}}{\partial t} + C . \overline{u}  + \nu  S . \overline{u} - F. \overline{u} = 0 ,
 \end{equation}
where $M$ , $C$, and $S$ are the mass, convection, and stiffness matrices respectively. $F$ can be obtained by linearizing
the source function $f$, and is thus a Jacobian matrix.
Thus, $M_{i, j} = \int_{\tau_h}   \lambda_i   \lambda_j  \,d\tau$ ,    \  $C_{i, j} =  \int_{\tau_h}  \lambda_i  \bm{V} \bm{\nabla} \lambda_j  \,d\tau $,
 and $S_{i, j} =  \int_{\tau_h}  \bm{\nabla}  \lambda_i   \bm{\nabla} \lambda_j  \,d\tau$. For the Burgers equation,  when implementing the system, the $V_x$
component used in $C$ matrix must be known (the value of $u_h$ at the previous time step can be used for example).

\subsection{The characteristic-Galerkin method}
It is well known that standard Galerkin finite element methods fail for advection and advection dominated advection-diffusion problems.
Among the different available stabilization techniques, we choose the method of characteristics  to discetize the Lagrangian derivative
$\frac{\partial  }{\partial t} + (\bm{V}\cdot\bm{\nabla}) $ in the two first equations  \citep{1992CMAME.100..117P}.
The finite elements Freefem++ software allows to do this in a simple way \citep{hec12}.
Indeed, the discretized version of the above Lagrangian derivative operator (for $u$) can be approximated at time $t_{n+1} = t_n + \Delta t$ 
($\Delta t$ being the time step) as,
\begin{equation}
\frac{\partial u}{\partial t} + (\bm{V}\cdot\bm{\nabla}) u = 
     \frac{u^{n+1}(x,y,t_{n+1}) - u^{n}\circ\left(\bm{X}^n(x,y,t_{n+1})\right)}{\Delta t_{}} ,
\end{equation}
where  $\bm{X}^n(x,y,t_{n+1})$ is the space location of a particle obtained by integrating backward (in time) along the characteristic curves from a node (at $t_{n+1}$).
Thus, $u^{n}\circ\left(\bm{X}^n(x,y,t_{n+1})\right)$ is the corresponding value obtained for $u$, that can be interpreted as particular value at $t_n$, ${u'}^{n}$.
In Freefem++, we use a first order approximation taking $\bm{V^n}$ to do this integration and thus deduce ${u'}^{n}$.
Moreover, the corresponding time discretized matrix formulation similar to Equation 6 can be automatically builded in FreeFem++ as,
\begin{equation}
    (\frac{M} {\Delta t} + \nu S - F) . \overline{u}^{n+1} = ( \frac{M} {\Delta t} - C^n) .  \overline{u}^n ,
 \end{equation}
where the diffusion is discretized fully implicitely, and $C^n$ is the convection operator using the flow velocity $\bm{V^n}$ \citep{hec12}.

The previous scheme is only first order in time, and its extension to a second order is not trivial \citep{rui02}.
We propose to use the following two-step predictor-scheme,
\begin{equation}
    (\frac{M} {\Delta t /2} + \nu S - F) . \overline{u}^{*} = ( \frac{M} {\Delta t /2} - C^n) .  \overline{u}^n ,
     \end{equation}
      \begin{equation}
    (\frac{M} {\Delta t} +  \frac{ \nu}  {2} S  - \frac{ F}  {2}) . \overline{u}^{n+1} = ( \frac{M} {\Delta t } - \frac{ \nu}  {2} S + \frac{ F}  {2} ) .   \overline{u}^n  - C^* .  \overline{u}^*  ,\\
 \end{equation}
where the predictor step gives a solution value $\overline{u}^{*}$ at $t + \Delta t /2$, and
allows to estimate a velocity value $V^*$ and a second order convection term with $C*$ operator for the corrector step. At the corrector
step, the viscous term also needs to be treated with a second order semi-implicit (Crank-Nicholson like) discretization in order
to get a second-order discretization.

Finally, a linear solver is chosen in order to solve the systems (Equation 14 and Equations 15-16). In Freefem++, a large choice of direct and iterative solvers
are available.

\subsection{The adaptive mesh}
In Freefem++, it is possible to adapt the mesh according to a giving function by using the Hessian matrix of this function.
In this way, the local error is made nearly uniform by choosing a non uniform mesh. This procedure is based on the
a posteriori error estimates, and is shown to be very efficient by choosing the solution at a given time itself for the function
(see the tests below).

\section{Tests on the simplified equations}

\subsection{Advection-diffusion of a Gaussian bell}
As a first example, we consider a Gaussian bell structure defined by,
\begin{equation}
u (x,y) =  \exp{ \left(-  \frac{(x - x_c)^2 + (y - y_c)^2}  {2 \sigma ^2 }  \right)}, 
 \end{equation}
of characteristic width $\sigma = 0.1$,
and initially centered around $(x_c = 0.5, y_c = 0)$ in a square domain $[-1 : 1]^2$. The bell is convected (according to Equation 9 with a zero source term $f = 0$)
with a given velocity field $\bm{V} = (-y, x)$ ( giving a solid rotation in clockwise direction), and is also diffusing with a viscosity coefficient $\nu$.
The exact solution at $t = t_f$ is,
\begin{equation}
 u (x,t) =     \frac{2  \sigma ^2 } {2  \sigma ^2  + 4  \nu t }  \exp{   \left(-  \frac{(x - x_f)^2  + (y - y_f)^2}  {2 \sigma ^2 + 4 \nu t }  \right) } .
 \end{equation}
We first apply the first order (in time) characteristic-Galerkin method using a fixed isotropic mesh (see Figure 1) with $P_1$ elements
and integrate during one turn (i.e. $t_f = 2 \pi$, $x_f = x_c$, $y_f = y_c$), using a rather small viscosity value $\nu = 1.5  \times 10^{-4}$ and a constant
time step $\Delta t = 0.023$.
The approximate solution on a crude mesh ($2116$
triangles) calculated at two different times (corresponding to a quarter of a turn, and one full turn) is plotted in Figure 2. A much better
approximation is obtained when using adapted meshes for a similar total number of triangles. Indeed, this is illustrated in Figure 3, where
the same test is done with an isotropic adapted mesh (i.e. with isogeometric triangles corresponding to a total number varying between $3200$ and
$2200$ at the end of the run), and also with a non-istropic adapted mesh (with a number of triangles varying between $2000$ and $1400$). The error is very similar
for these two adapted meshes runs, but the CPU time is advantageous when using non isogeometric elements as the number of triangles
is smaller. Note that for this test, the mesh adaptation is done at each time-step.
In Figure 4, we have plotted the $L^2$ error norm  as a function of the number of triangles $n_t$.
The time step  $\Delta t$ is also taken to vary like $ 1/\sqrt {n_t}$ in order to check the linear dependence expected of the $L^2$ error norm
on the solution with the mesh size $h$ of the triangles, as we employ the first order scheme.
Moreover, the results obtained by using our adaptive procedure with $P_1$ elements on the same test, lead to an error that is substantially lower
(divided by two approximately) for the same number of triangles (of order $2000$). Note that, an isotropic mesh is imposed for this test (see Figure 3),
and when a non-isotropic mesh is used the same error is obtained with an even smaller number of triangles ($1400$ versus $2200$).
The results are even further ameliorated when using $P_2$ elements instead, as one can see in Figure 4, as 
the error is divided by a factor of $5$ compared to a uniform mesh.
A constraint on the maximum edge size $h_{max} = 0.5$ is used, that is not fundamental for our conclusions.

In summary, this means that for a given accuracy, a gain factor of order $100$ in the number of elements is obtained thanks
to the adaptive scheme with $P_2$ elements, comparatively to a uniform mesh calculation with $P_1$ elements. The CPU time
is correlatively much less important. Finally note that, the choice of the solver is not fundamental but is slightly better for CPU time for this test when using a direct
one (UMFPACK) compared to an iterative one (GMRES).

\subsection{Unsteady advection-diffusion with boundary and internal layers}
We present a challenging test where an unsteady strongly anisotropic solution for the advection equation (Equation 9)
with zero source term $f = 0$ \citep{2018CMAME.340..864M}. Indeed, we set an
condition $u_0 (x, y) = 0$ in the domain $[0: 1]^2$ except on the boundary where,
\begin{equation}
 u_0 (x , y) =
  \left\{
    \begin{aligned}
    1 \ \ \ \ \ \ \ \ \ \ if \ \ x = 0, \ \ 0   \le y \le 1 ,\\
    1 \ \ \ \ \ \ \ \ \ \ if \ \  0   \le x \le 1, \ \ y = 1 ,\\
    (\delta - x)/\delta \ \ \ \ \ \ \ \ \ \ if \ \ x \le \delta, \ \ y = 0  ,\\
    0 \ \ \ \ \ \ \ \ \ \ if \ \ x > \delta, \ \ y = 0  ,\\
     (y -1 + \delta)/\delta \ \ \ \ \ \ \ \ \ \ if \ \ x = 1, \ \ y  \ge 1- \delta  ,\\
     0 \ \ \ \ \ \ \ \ \ \ if \ \ x = 1, \ \ y  < 1- \delta , \\
      \end{aligned}
   \right.
 \end{equation}
that is also kept during time evolution. A constant velocity field $\bm{V} = (2, 1)$ is taken with a viscosity coefficient $\nu = 10^{-3}$.
We also use $ \delta = 7.8125  \times10^{-3}$. Thus, this problems exhibits boundary layers along $x = 0$ and $y = 1$.
A time step $\Delta t = 0.001$ is taken to integrate the equation with our first-order (in time) characteristic-Galerkin method
using the efficient scheme with triangles ($P_2$ elements), as described for the previous test.

As time advances, the left boundary layer propagates into the domain creating an internal layer which finally reaches the right boundary
and creating a new boundary layer because of the imposed $u$ value. The initial top boundary layer also reduces progressively in the
same time (as illustrated in Figures 5-7). The numerical solution obtained at different times is clearly well captured by our anisotropic adapted
mesh, and the quality of our results is similar to the best ever developed schemes \citep{2018CMAME.340..864M}.
Note that, as mentioned
in the previously cited study and references therein, a fixed mesh is unable to capture the layers without strong oscillations unless a prohibitive number of triangles is used.
The number of triangles remains relatively low with our scheme, as it varies between $2800$ (at the end) and $12000$ (at early time) in this
case, with no constraint on isotropy but an arbitrary constraint on the maximum edge size $h_{max}  = 0.4$.

\subsection{Viscous shock layer induced with Burgers equation}
A third (and last) test is proposed in order to check the ability of our method to capture the formation of shock structure. Indeed, we
consider now the solution of the viscous Burgers equation (Equation 10) with a zero source term $f = 0$ and a velocity component $V_y = 1$.
Following the initial setup \citep{che13},
we set an initial  condition $u_0 (x, y) = 1.5 - 2x$ in a $[0: 1]^2$ domain,
except at the bottom boundary (at $y = 0$) where a Dirichlet boundary value $u = 1.5 - 2x$ is taken and kept constant in time.
An extremely challenging low viscosity value is also used, as $\nu = 1 \times 10^{-5}$.
The results obtained with our adaptive method (same scheme as used for the above previous test problem
using $P_2$ elements and an adaptive procedure applied at every time step) with a constant time step $\Delta t = 0.004$ are presented in Figures 8-9.
A shock layer is formed, beginning at ($x = 0.75$, $y = 0.5$) as a consequence of the steepening of the slope solution when propagating 
upward and rightward, that has been initiated at the bottom layer. As time advances, a final steady-state solution having an oblique shock layer
between the previous point and the top-right corner is obtained, as one can see on Figure 8 at $t = 11.5 $.
The final state structure showing the characteristics curves is nicely reproduced in Figure 9.
Depending on time, the number of triangles in our adapted mesh scheme varies between
$500$ and $17000$ in this case, with no constraint on isotropy but constraint on the maximum size $h_{max} = 0.3$.

In summary, our scheme nicely reproduces the shock-involving solution despite our scheme is not conservative.

\begin{figure}
\centering
 \includegraphics[scale=0.37]{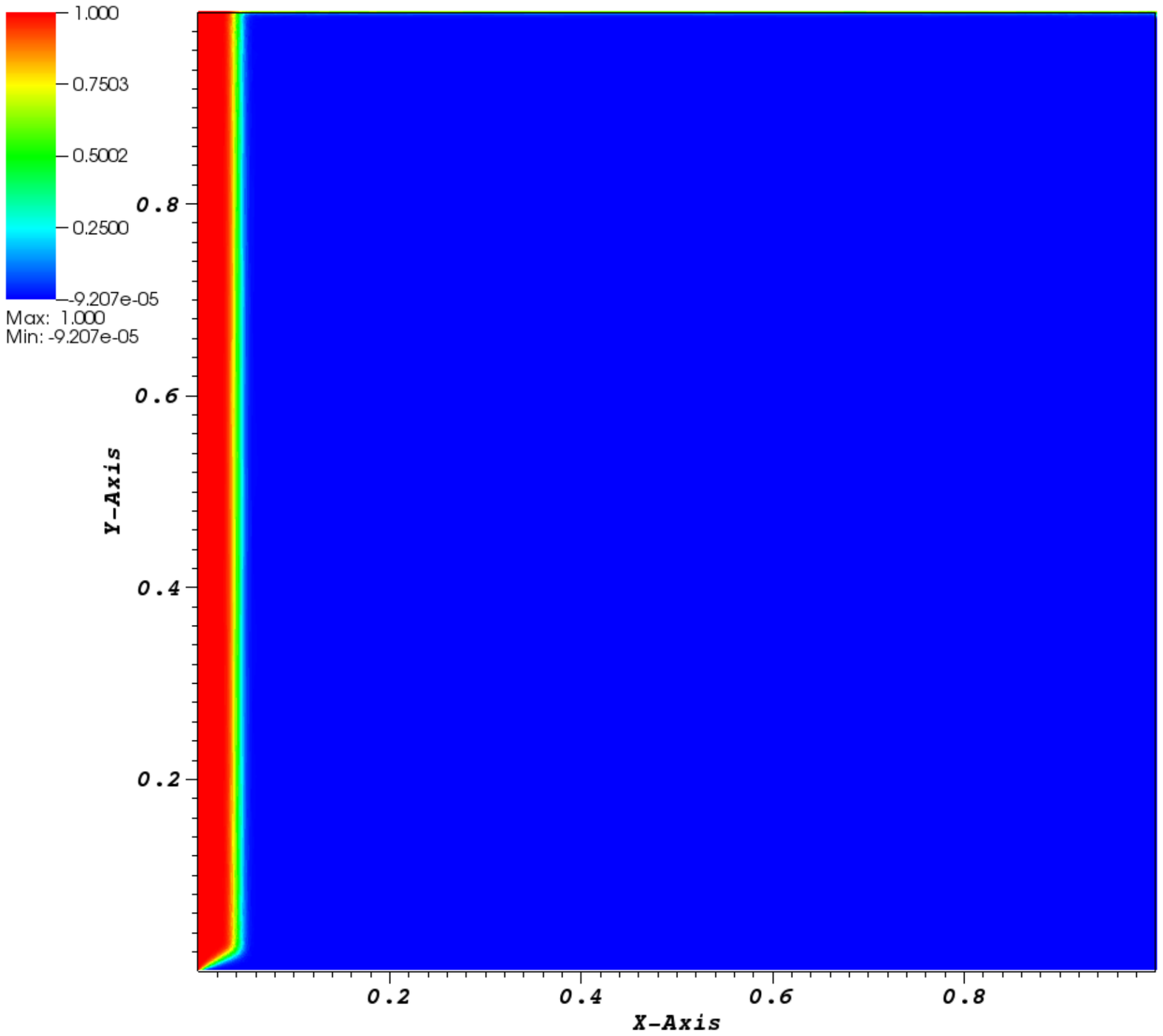}
 \includegraphics[scale=0.37]{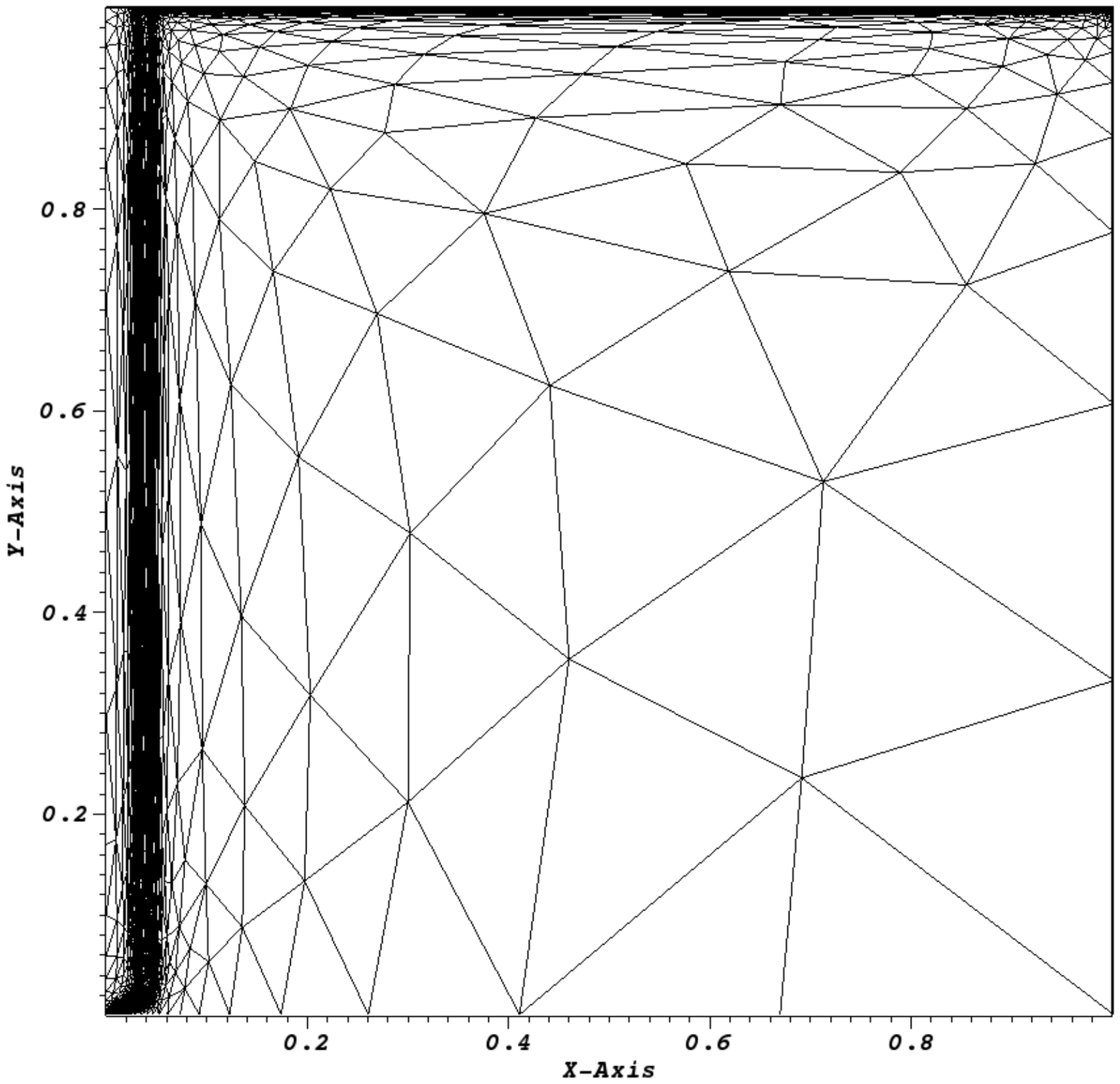}
  \caption{Contour map solution (left panel) of the unsteady advection diffusion problem (see text) obtained at $t  = 0.02$, and
  corresponding adaptive mesh (right panel).}
\label{fig5}
\end{figure}

\begin{figure}
\centering
 \includegraphics[scale=0.37]{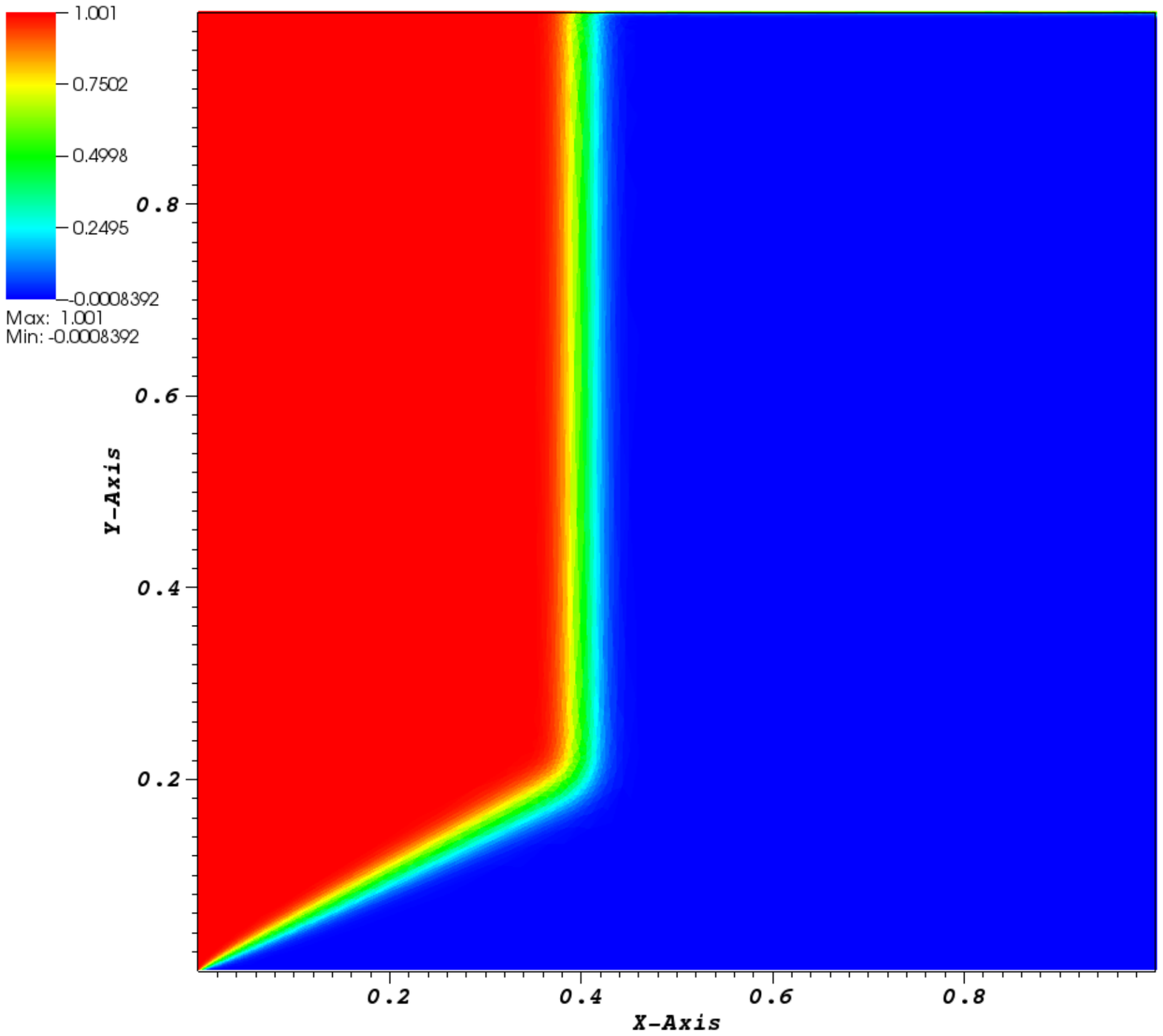}
 \includegraphics[scale=0.37]{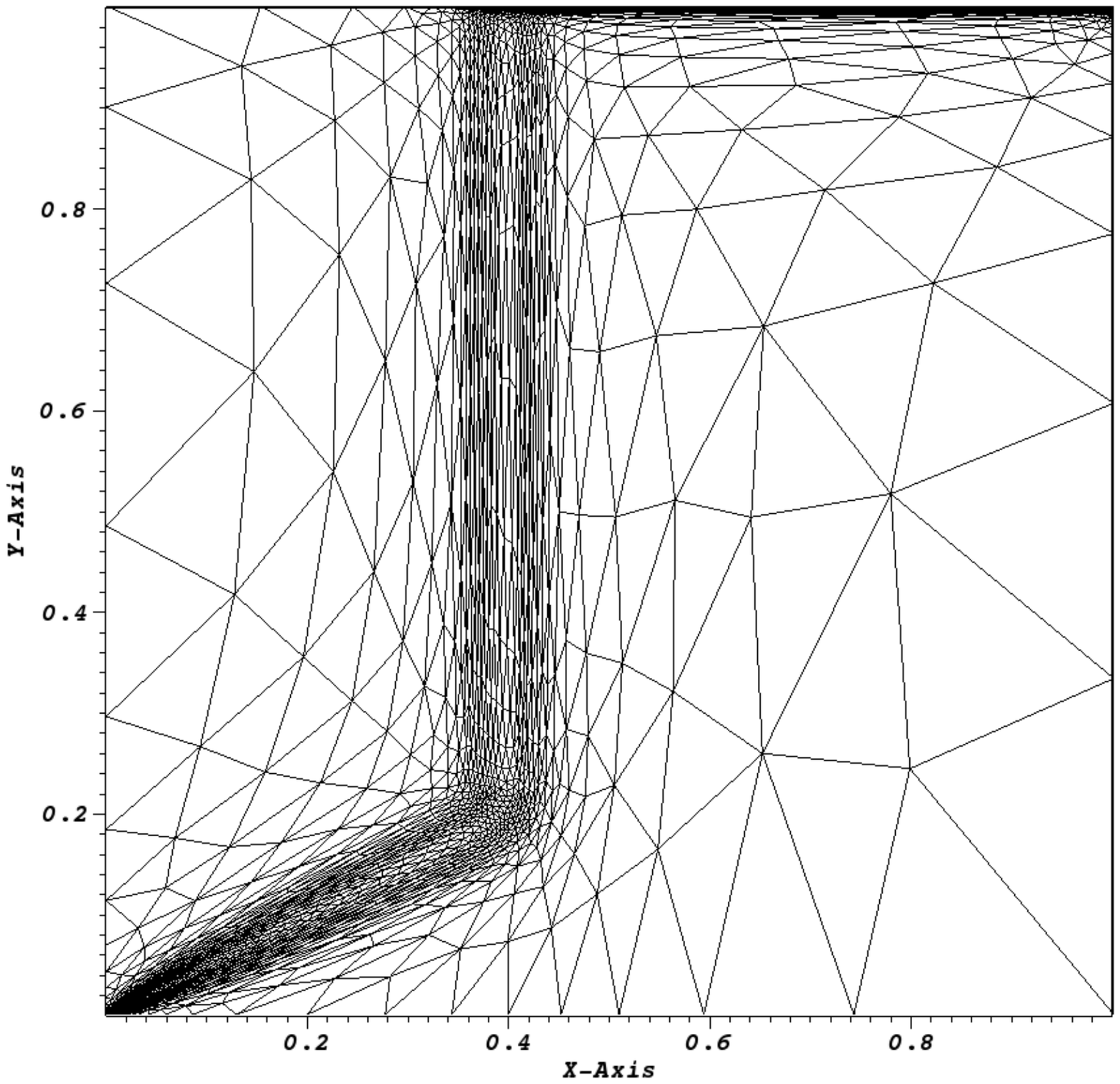}
  \caption{Same as previous figure for $t = 0.2$.}
\label{fig6}
\end{figure}

\begin{figure}
\centering
 \includegraphics[scale=0.37]{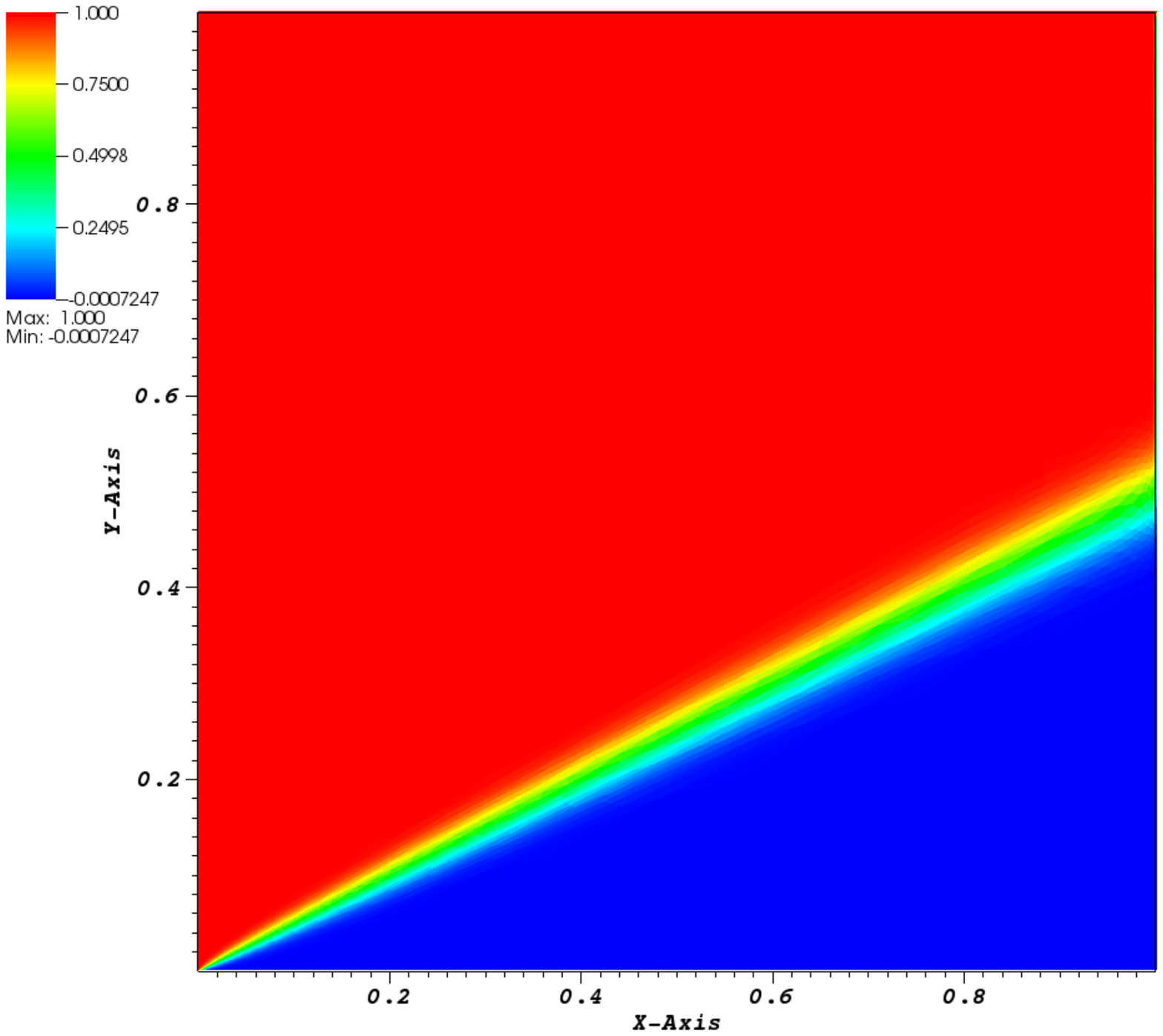}
 \includegraphics[scale=0.37]{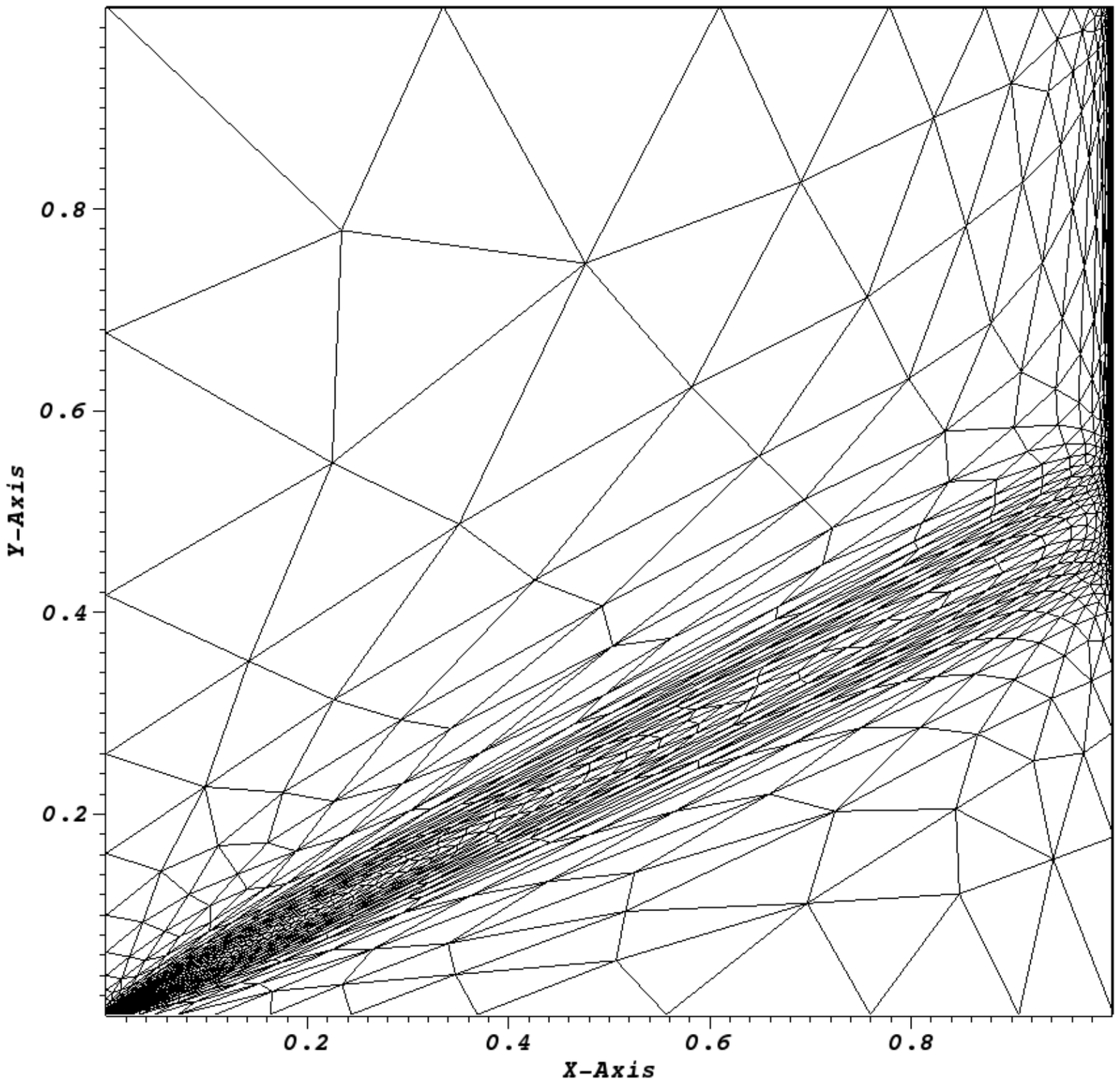}
  \caption{Same as previous figure for $t = 0.6$.}
\label{fig7}
\end{figure}

\begin{figure}
\centering
 \includegraphics[scale=0.37]{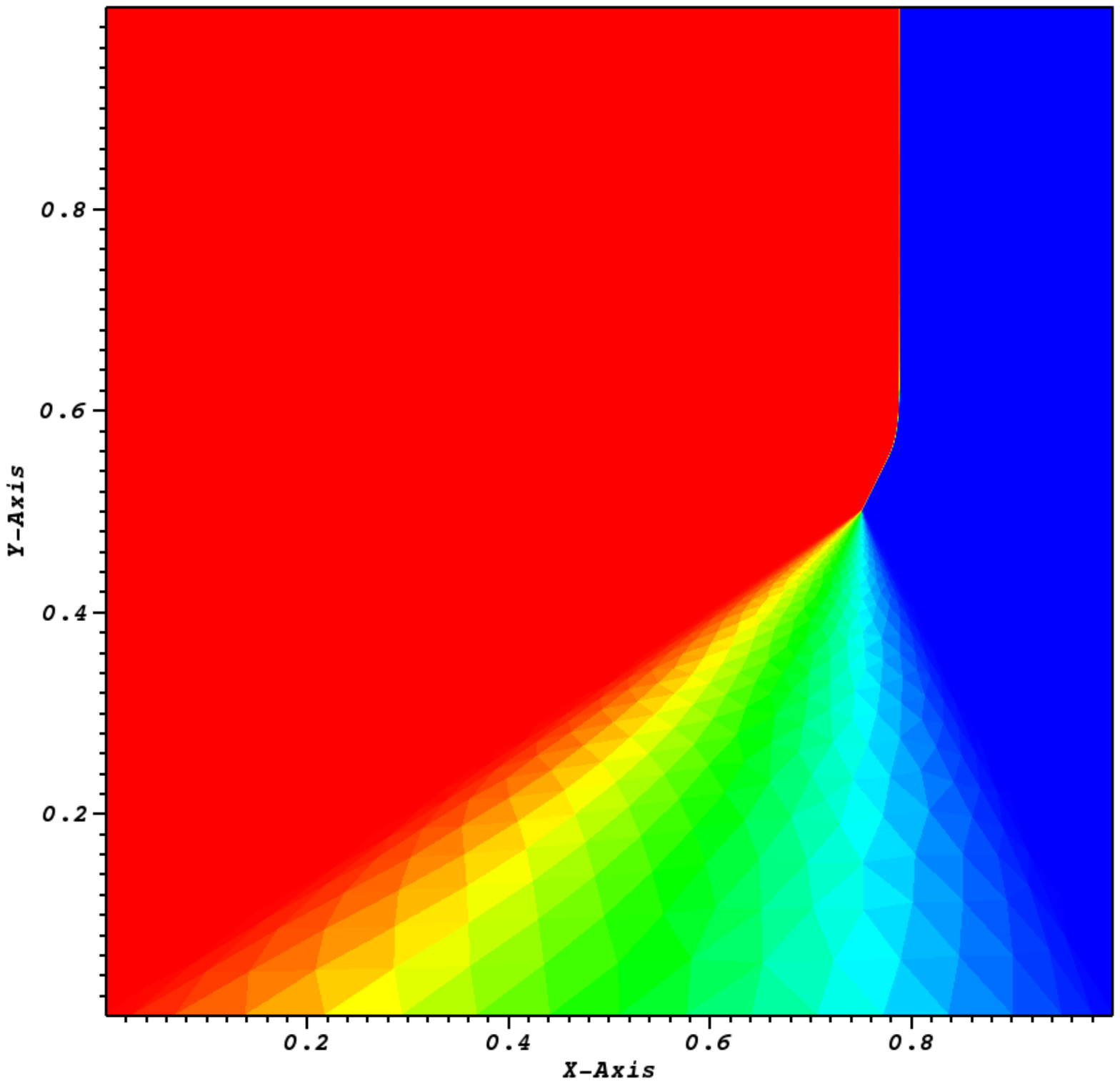}
 \includegraphics[scale=0.37]{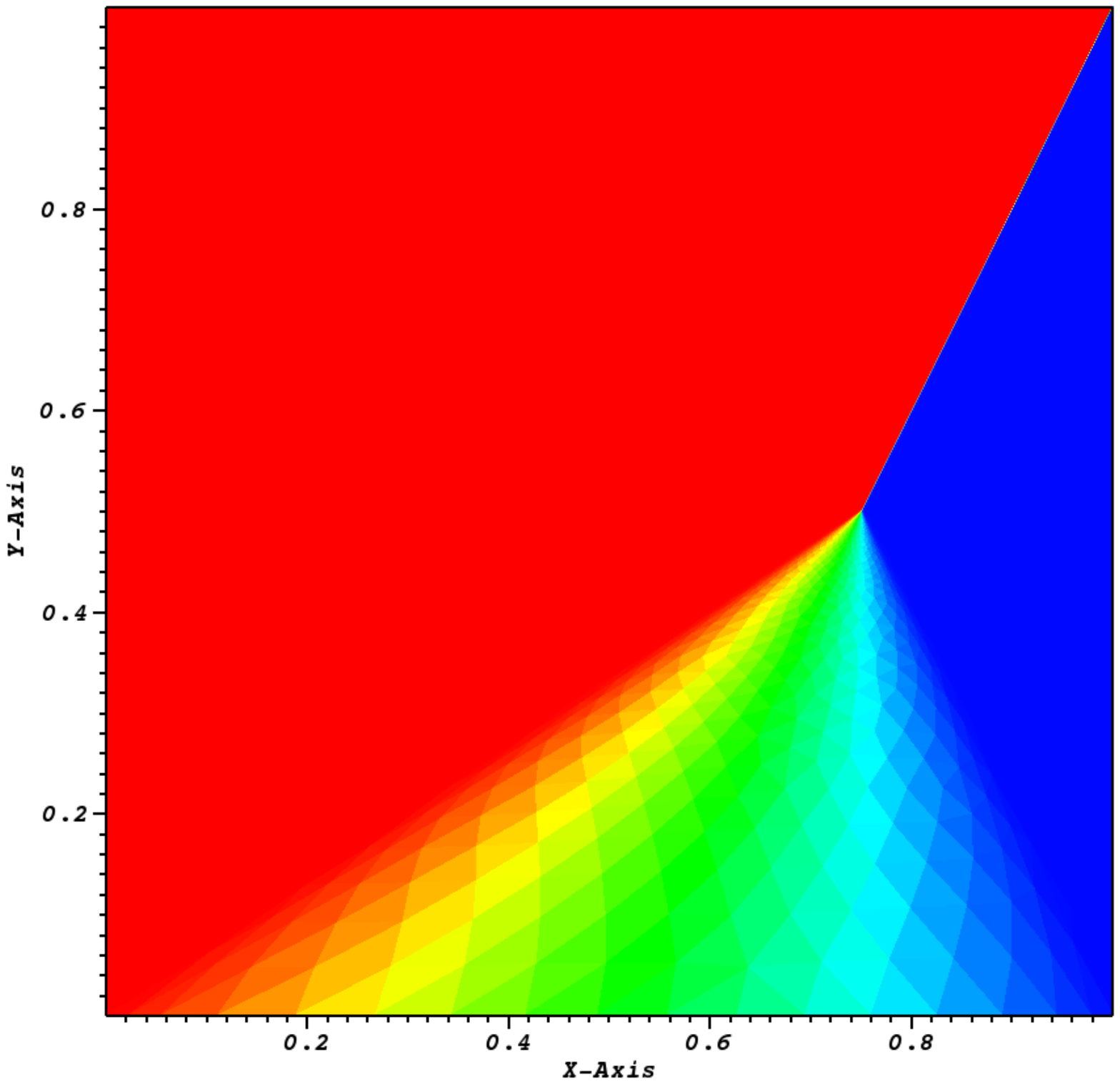}
  \caption{Contour map solution of the viscous shock layer problem (see text) obtained at two times, $t = 1.5$ (left panel)
   and the final time $t = 11.5$ (approximate steady state solution, right panel).}
\label{fig8}
\end{figure}

\begin{figure}
\centering
 \includegraphics[scale=0.37]{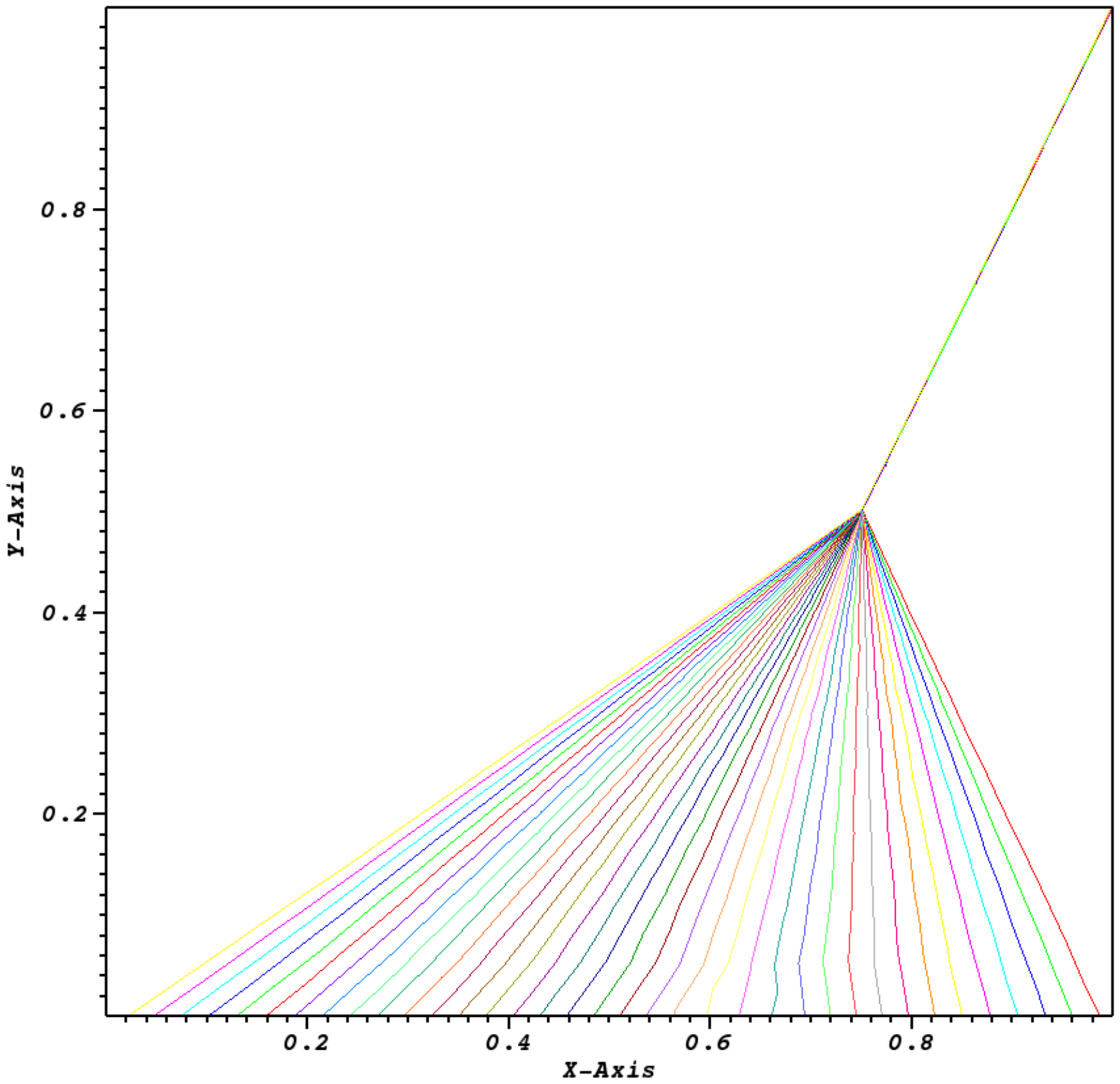}
 \includegraphics[scale=0.37]{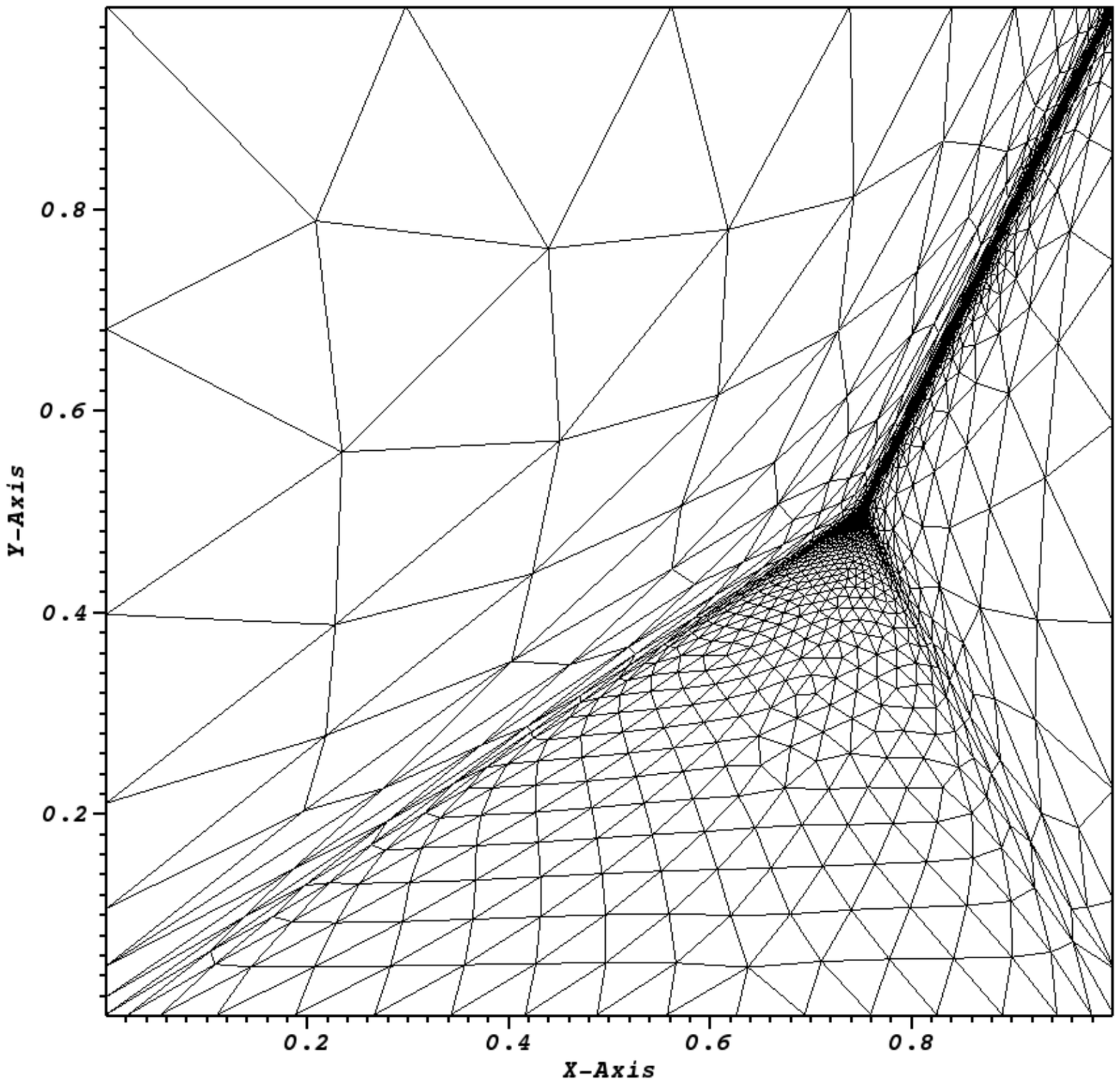}
  \caption{Final steady state solution (see previous figure) using $36$ iso-contour lines (left panel), and corresponding mesh grid (right panel).}
\label{fig9}
\end{figure}

\section{FINMHD code for tilt instability and magnetic reconnection}
Our characteristic-Galerkin method is now applied to the 2D set of  reduced MHD equations written in current-vorticity variables (Equations 5-8).

\subsection{FINMHD code}
We apply the first-order scheme (Equation 14) to the MHD equations. This is basically a non-linear problem, because of the coupling terms
in the different equations, that needs to be solved using linear solvers. Different options are possible. We choose the
following scheme, where the two (time evolution) equations are solved as a single system as,
\begin{equation}
    \left\{
    \begin{aligned}       
          \frac{\omega^{n+1} - \omega^{n}\circ X^n}{\Delta t_{}}  =    (\bm{B^n}\cdot\bm{\nabla})J^{n+1} + \nu \bm{\nabla}^2 \omega^{n+1} ,     \\    
                 \frac{J^{n+1} - J^{n}\circ X^n}{\Delta t_{}}  =    (\bm{B^n}\cdot\bm{\nabla})\omega^{n+1} + \eta \bm{\nabla}^2 J^{n+1}   + g(\psi^n,\phi^n)  ,\\ 
    \end{aligned}
    \right.
\end{equation}
where the two diffusive termes are treated fully implicitly, and the non linear g term fully explicitly.
The two last equations are successively integrated in a separate way, once the solutions for $\omega^{n+1}$ and $j^{n+1}$ are obtained, as
\begin{equation}
    \begin{aligned}    
        \bm{\nabla}^2 \phi^{n+1}   = - \omega^{n+1} ,\\
    \end{aligned}
\end{equation}
\begin{equation}
    \begin{aligned}    
        \bm{\nabla}^2 \psi^{n+1}   =  - J^{n+1} .\\
    \end{aligned}
\end{equation}
The above scheme is only first order in time, but has been shown to be linearly unconditionally stable (see Appendix A).

A second-order time integrator using a predictor-corrector scheme has also been developed following the idea proposed previously (Equations 15-16),
\begin{equation}
    \left\{
    \begin{aligned}
             \frac{\omega^{*} - \omega^{n}\circ X^n}{\Delta t_{}/2}  =    (\bm{B^n}\cdot\bm{\nabla})J^{*} + \nu \bm{\nabla}^2 \omega^{*}    ,  \\         
                \frac{J^{*} - J^{n}\circ X^n}{\Delta t_{}/2}  =    (\bm{B^n}\cdot\bm{\nabla})\omega^{n} + \eta \bm{\nabla}^2 J^{*}   + g(\psi^n,\phi^n)  ,\\       
    \end{aligned}
    \right.
\end{equation}
and
\begin{equation}
    \begin{aligned}    
        \bm{\nabla}^2 \phi^{*}   = - \omega^{*} ,\\
    \end{aligned}
\end{equation}
\begin{equation}
    \begin{aligned}    
        \bm{\nabla}^2 \psi^{*}   =  - J^{*} ,\\
    \end{aligned}
    \end{equation}
for the predictor step. One must note that the term $\bm{B^n}\cdot\bm{\nabla} $ is evaluated for $\omega^{n}$ and not for $\omega^{*}$,
as it is known to lead to better stability property. The corrector step follows,
\begin{equation}
    \left\{
    \begin{aligned}
             \frac{\omega^{n+1} - \omega^{n}\circ X^*}{\Delta t_{}}  =    (\bm{B^*}\cdot\bm{\nabla})J^{*} + \frac{\nu}  {2} ( \bm{\nabla}^2 \omega^{n+1} 
          +   \bm{\nabla}^2 \omega^{n} )  , \\         
          \frac{J^{n+1} - J^{n}\circ X^*}{\Delta t_{}}  =    (\bm{B^*}\cdot\bm{\nabla})\omega^{*} + \frac{\eta}  {2} ( \bm{\nabla}^2 J^{n+1}
           +  \bm{\nabla}^2 J^{n} )  + g(\psi^{*} ,\phi^*) , \\     
     \end{aligned}
    \right.
\end{equation}
and 
 \begin{equation}
    \begin{aligned}    
        \bm{\nabla}^2 \phi^{n+1}   = - \omega^{n+1} ,\\
    \end{aligned}
\end{equation}
\begin{equation}
    \begin{aligned}    
        \bm{\nabla}^2 \psi^{n+1}   =  - J^{n+1} .\\
    \end{aligned}
\end{equation}
Contrary to the first-order scheme, our second-order variant is linearly conditionally stable. Moreover, the CFL limitation
(see Appendix A) requires an initial time step value that is one order of magnitude smaller than the value required for
converged solution obtained by the first-order scheme. This has been numerically checked, leading to the choice of using the
first-order scheme in the following magnetic reconnection study.

\subsection{Magnetic reconnection using FINMHD}

 \subsubsection{Equilibrium configuration and numerical setup for tilt instability}
The initial magnetic field configuration for tilt instability is a dipole current structure similar to the dipole vortex flow pattern in
fluid dynamics \citep{1990PhFlB...2..488R}.
It consists of two oppositely directed currents embedded in a constant magnetic field (see Figure 10).
Contrary to the coalescence instability based on attracting current structures, the two antiparallel currents in the configuration tend to repel.
The initial equilibrium is thus defined by taking the following magnetic flux distribution,
\begin{equation}
    \psi_e (x, y)=
    \left\{
      \begin{aligned}
        &\left(\frac{1}{r} - r\right)\frac{y}{r} ~~~& & if ~~ r > 1 , \\
        &-\frac{2}{kJ_0(k)}J_1(kr)\frac{y}{r} ~~~& & if ~~ r\leq1 .\\
      \end{aligned}
      \right.
  \end{equation}
  
 And the corresponding density current is,
       \begin{equation} 
    J_e (x, y) =
    \left\{
      \begin{aligned}
        &~~~~~~~~~~~~0 ~~~& & if ~~ r > 1 , \\
        &-\frac{2k^2}{kJ_0(k)}J_1(kr)\frac{y}{r} ~~~& & if ~~ r\leq1 ,\\
      \end{aligned}
    \right.
\end{equation}

\noindent where 
$r=\sqrt{x^2+y^2}$, and $J_0$ et $J_1$  are Bessel functions of order $0$ and $1$ respectively.
Note also that $k$ is the first (non zero) root of $J_1$, i.e. $k = 3.83170597$.
\medskip
This initial setup is similar to the one used in the previously cited references \citep{1990PhFlB...2..488R, 2007JCoPh.225..363L},
and rotated with an angle of $\pi/2$ compared to the
equilibrium chosen in the other studies \citep{2014ApJ...795...77K, 2017MNRAS.467.3279R}.
Note that, the asymptotic (at large $r$) magnetic field strength 
is unity, and thus defines our normalisation. Consequently, our unit time in the following paper, will be defined as the Alfv\'en transit
time across the unit distance (i.e. the initial characteristic length scale of the dipole structure).
In usual MHD framework using the flow velocity and magnetic variables, force-free equilibria using an additional
vertical (perpendicular to the $x-y$ plane) can be considered \citep{1990PhFlB...2..488R}, or non force-free equilibria
can be also ensured trough a a thermal pressure gradient balancing the Lorentz force \citep{2014ApJ...795...77K}.
In our incompressible reduced MHD model, as thermal pressure is naturally absent, we are not
concerned by such choice. 

A stability analysis in the reduced MHD approximation using the energy principle has given that the linear
eigenfunction of the tilt mode is a combination of rotation and outward displacement \citep{1990PhFlB...2..488R}. Instead of
imposing such function in order to perturb the initial setup, we have chosen to let the instability develops from
the numerical noise. Consequently, an initial zero stream function is assumed $ \psi_e (x, y) = 0$,
with zero initial vorticity $ \omega_e (x, y) = 0$. Following most of the previous cited studies, this setup and the following computations are done in
a square domain $[-3: 3]^2$. The boundaries are taken to be relatively far from the centre, in order to have a weak effect on
the central dynamics. The initial current structure with a few magnetic field lines are reported
in Figure 10, with the initial mesh that is adapted using the Hessian matrix of the current density distribution. This arbitrary choice of current density is
justified by the fact that, first the initial dynamics of the tilt instability is driven by the current distribution (i.e. this is an ideal current driven MHD
instability), and second the ensuing magnetic process is controlled by the structure of the current layers.

\subsubsection{Typical evolution of the tilt instability and Sweet-Parker reconnection}
First, we focus on moderately low values of the resistivity and viscosity, and for simplicity we also assume a fixed Prandtl number,
$P_r = \nu / \eta = 1$. The early time evolution of the system corresponding to the tilt instability is well documented (see the previously cited studies).
It corresponds to the linear stability analysis, where the pair of oppositely directed
currents tend to repel one another giving rise to a rotation (see Figure 11 for a case with $\eta = 0.0025$) during the linear phase. The sense of rotation
(clock sense in Figure 11) is not predetermined and depends only on the numerical noise.
This rotation causes two new regions of enhanced current density at the leading edges of the vortices because of an associated outward
component of the linear displacement, taking the form of two bananas.
The time history of the maximum current density amplitude (taken over the whole domain) is shown to increase exponentially in time
over the equilibrium value (that is equal to approximately $10$), as illustrated in Figure 12 for three different runs using $\eta = 0.0025, 0.001$, and $0.0005$.
We have evaluated the corresponding local Lundquist number at saturation, $S = l V_A/\eta$, where $l$ is the full length
of each current layer and $V_A$ is the Alfv\'en velocity based on the magnetic field measured front of the layer. For example,
we have estimated that $l = 1.72$ and $V_A = 1.55$ for the case $\eta = 2.5  \times 10^{-3}$, leading to $S  \simeq 1080$. The values
of $l$ and $V_A$ are checked to slightly increase when the resistivity is decreased for $ \eta \simgt 0.0005$, and they become
approximately constant for smaller resistivity values.

A steady-state is obtained with a nearly constant current density structure subsequently forming, driving thus the reconnection process.
For example, an average value of $43$ is evaluated for the saturated current density in run using  $\eta = 2.5  \times 10^{-3}$.
The choice of the optimal parameters (as the initial time step and the adaptive criterion for its current value) necessary for the convergence of the results is reported in Annexe B.
For example, the time scale $\tau = (2.6)^{-1} \simeq 0.38 $ from the current density growth rate is shown to be well reproduced, independently of the
resistivity as the tilt mode is an ideal instability. The oscillations in the current density around the average value are due to the sloshing phenomenon,
as described in the coalescence problem between two magnetic islands \citep{2006PhPl...13c2307K}, because of the magnetic pressure buildup effect in
thin current sheets. The detailed structure of the current sheets structure, the adapted mesh, and corresponding magnetic field lines are visible on
Figure 13. Indeed, one can clearly see in right panel of Figure 13, the few tens of elements covering the width of the current layers.
The beginning of the magnetic reconnection mechanism is also well illustrated in left panel of Figure 13 for a time corresponding to the early
saturation of the current density.
The characteristic reconnection time $\tau_r$ can be determined in different equivalent ways. The simplest one is to take the whole duration of the process,
and it corresponds to the elapsed time between the saturation (first peak) and the
final time when the current density in Figure 12 returns to a very small value corresponding to a new state
(free of closed magnetic field lines). A second way is to measure the level of maximum current amplitude $J_{max}$  during reconnection (horizontal
lines in Figure 12), and to multiply it by the resistivity, as $\eta J_{max}$ is a measure of the reconnection rate (inverse of $\tau_r$) for a steady-state
process. We have checked that the reconnection rate for $S  \simlt 10^4$
closely follows a SP scaling, as $J_{max} \propto S^{1/2}$ (see next subsection).
  
In the run employing $\eta = 0.0005$, i.e. corresponding to a local Lundquist number $S  \simeq 8000$, after two oscillations of the maximum
current density at saturation (i.e. $t  \simeq 10.7$), a single plasmoid has been observed to form in a transient way (see Figure 14).
This plasmoid does not influence the value of the saturated current density and the resulting reconnection rate.
This indicates the proximity of the critical Lundquist number as $S_c  \simeq 10^4$. The exact value of the critical Lundquist 
depends on the numerical noise \citep{2017ApJ...849...75H}.
  
 \begin{figure}
\centering
 \includegraphics[scale=0.37]{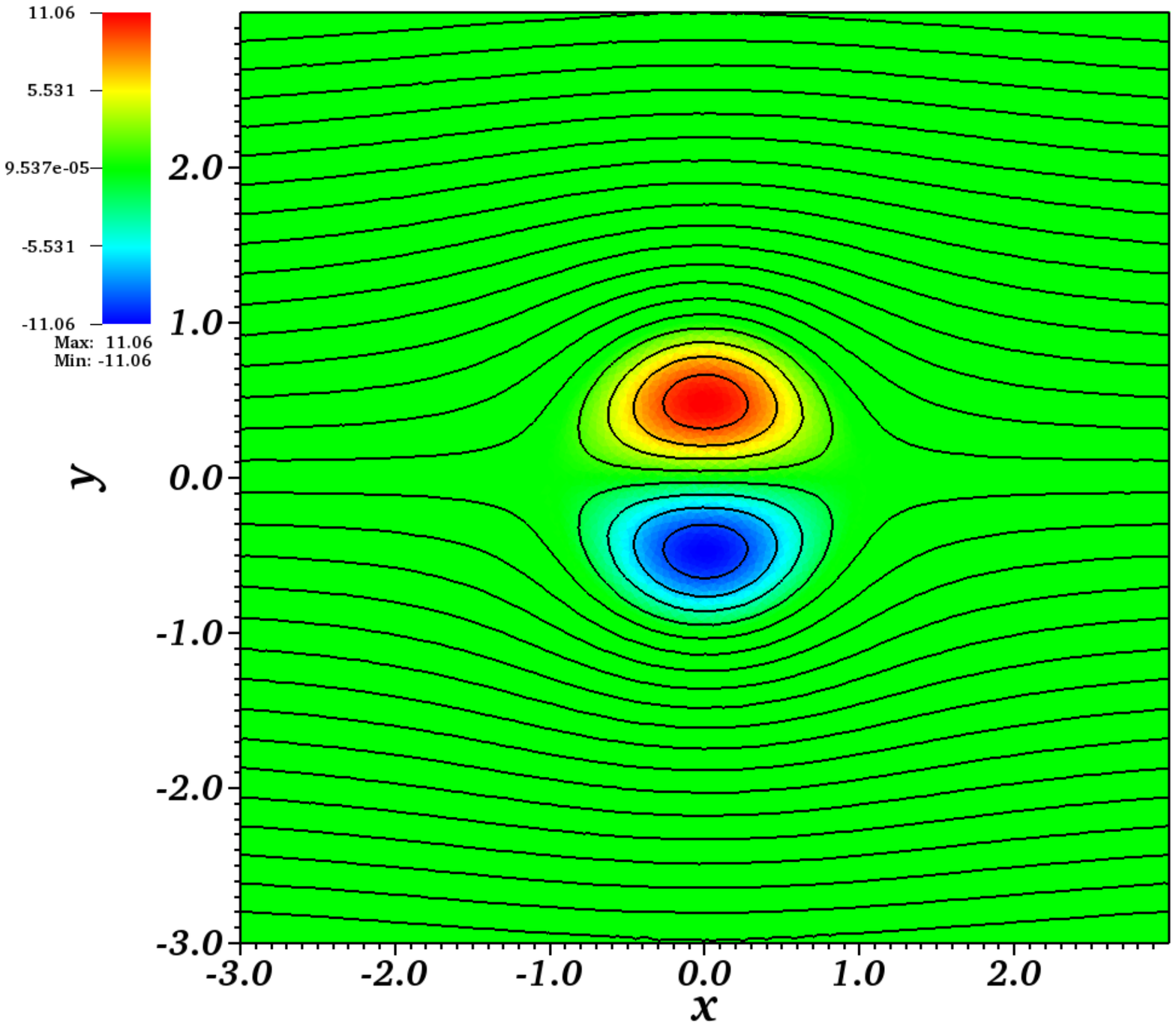}
 \includegraphics[scale=0.37]{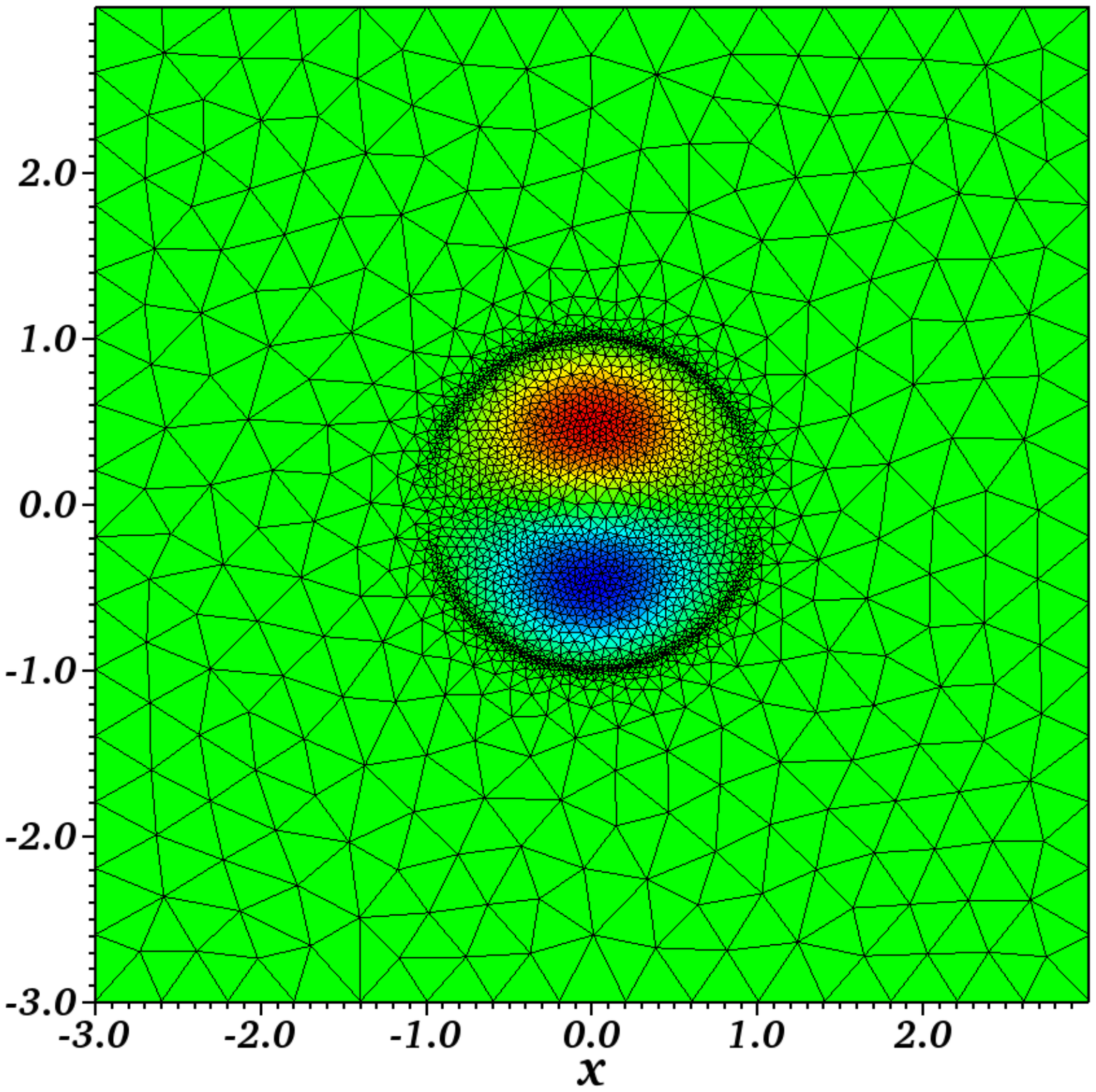}
  \caption{Initial configuration for the tilt instability showing the dipole current density structure overlaid with magnetic field lines (left panel), and overlaid 
  with the initial grid using the density current to adapt the mesh (right panel).
  }
\end{figure}
 
 \begin{figure}
\centering
 \includegraphics[scale=0.37]{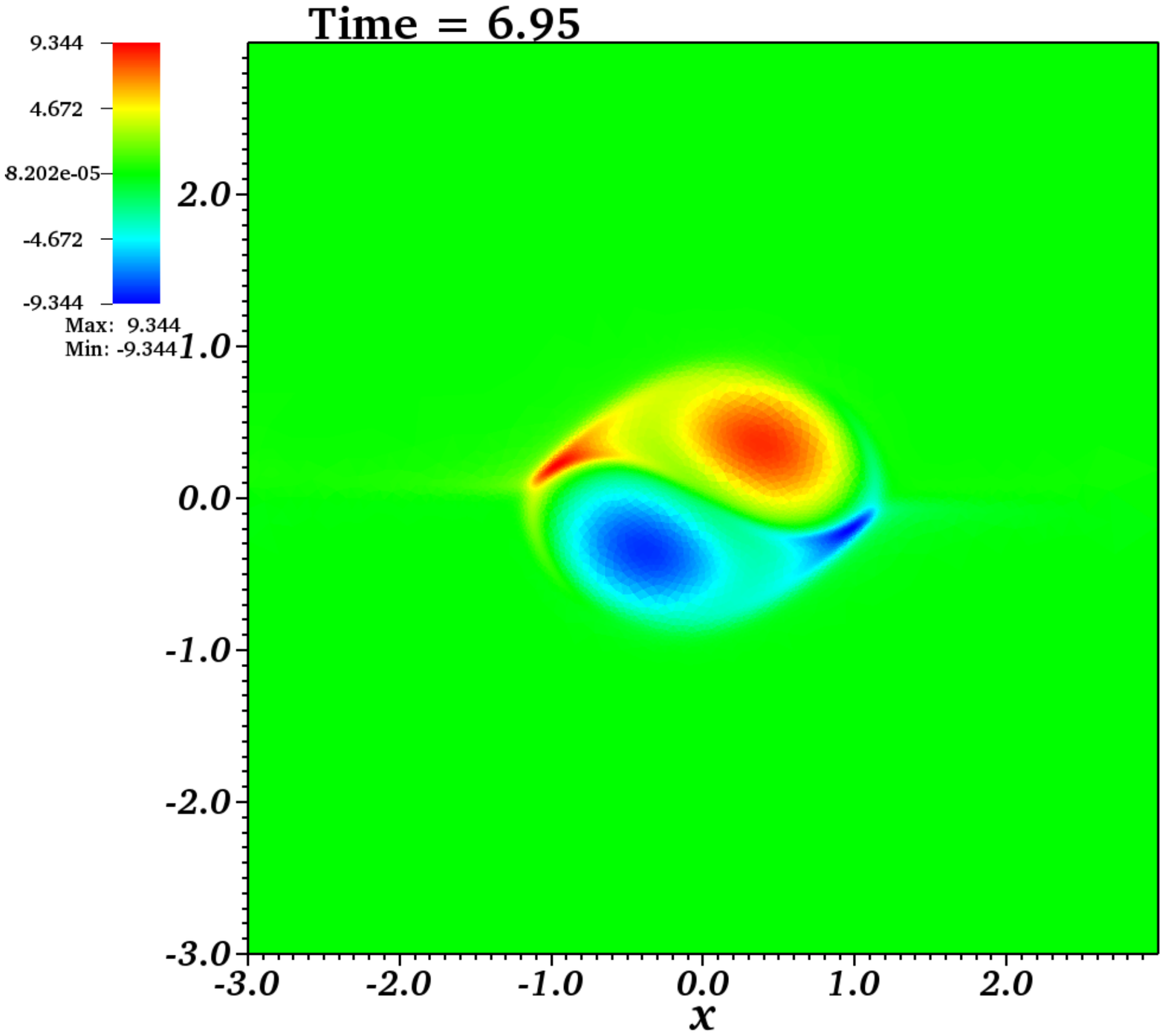}
 \includegraphics[scale=0.37]{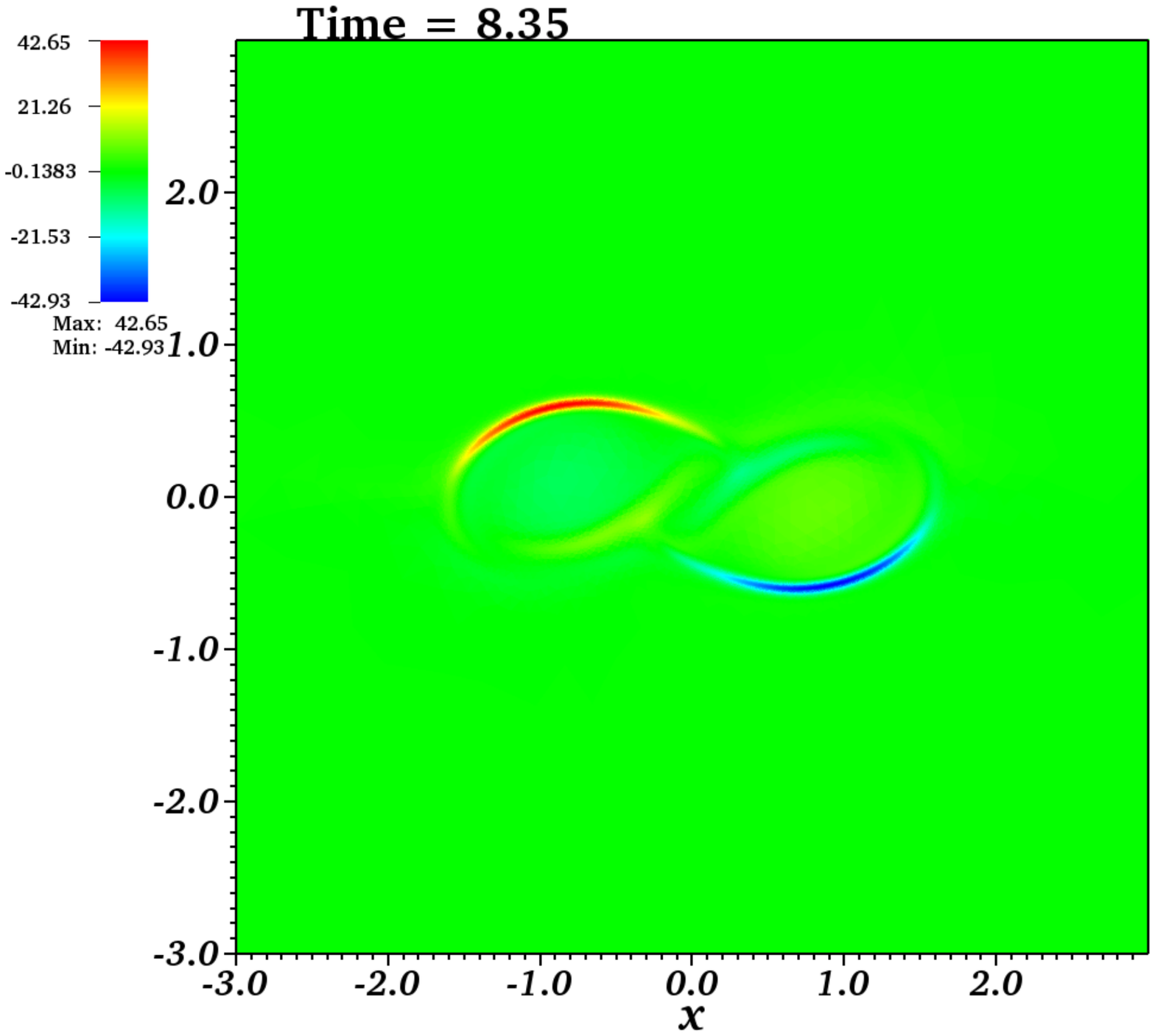}
  \caption{Current density at different times during the development of the tilt instability for $\eta = 0.0025$, $t = 6.95$ (left panel), and $t = 8.35$ (right panel).
  }
\end{figure}

 \begin{figure}
\centering
\includegraphics[scale=0.5]{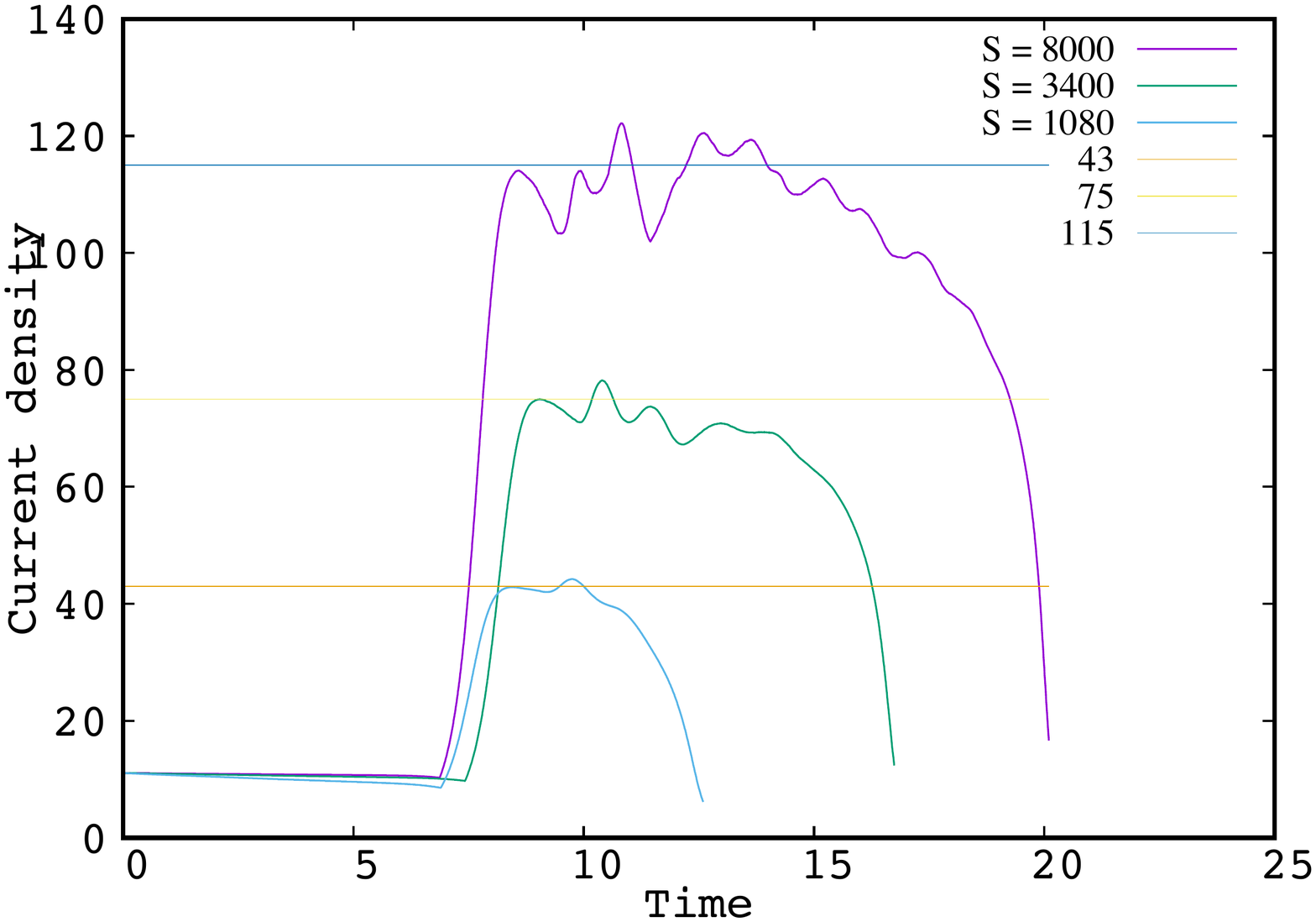}
  \caption{Time history of the maximum
  current density amplitude for three resistivity values (i.e. $\eta = 0.0025, 0.001$, and $0.0005$) corresponding to the
 estimated Lundquist number values, $S  \simeq 1080, 3400, 8000$ respectively. The horizontal lines indicate the values for
 the average saturated current densities.
    }
\end{figure}

 \begin{figure}
\centering
 \includegraphics[scale=0.37]{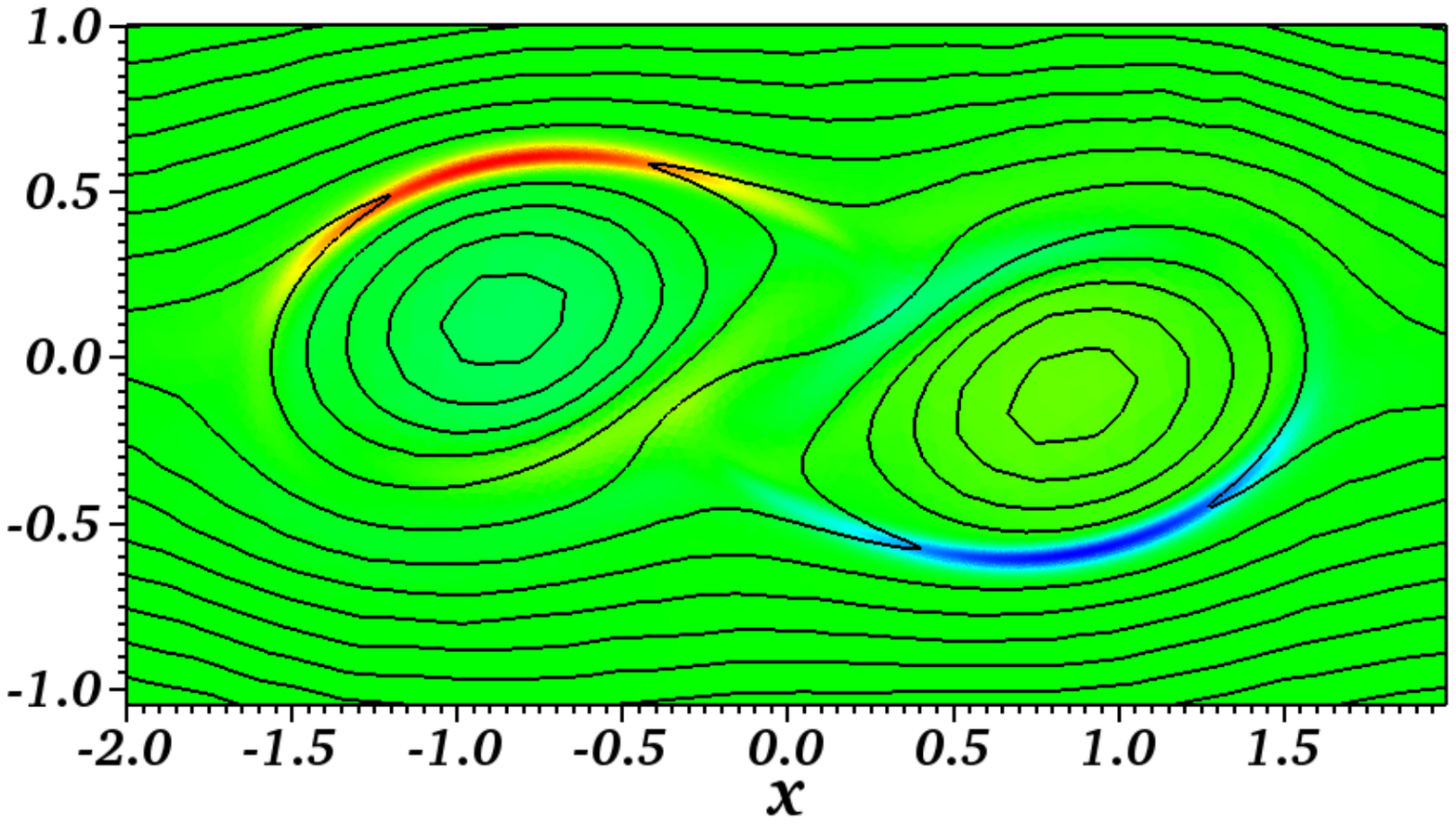}
 \includegraphics[scale=0.37]{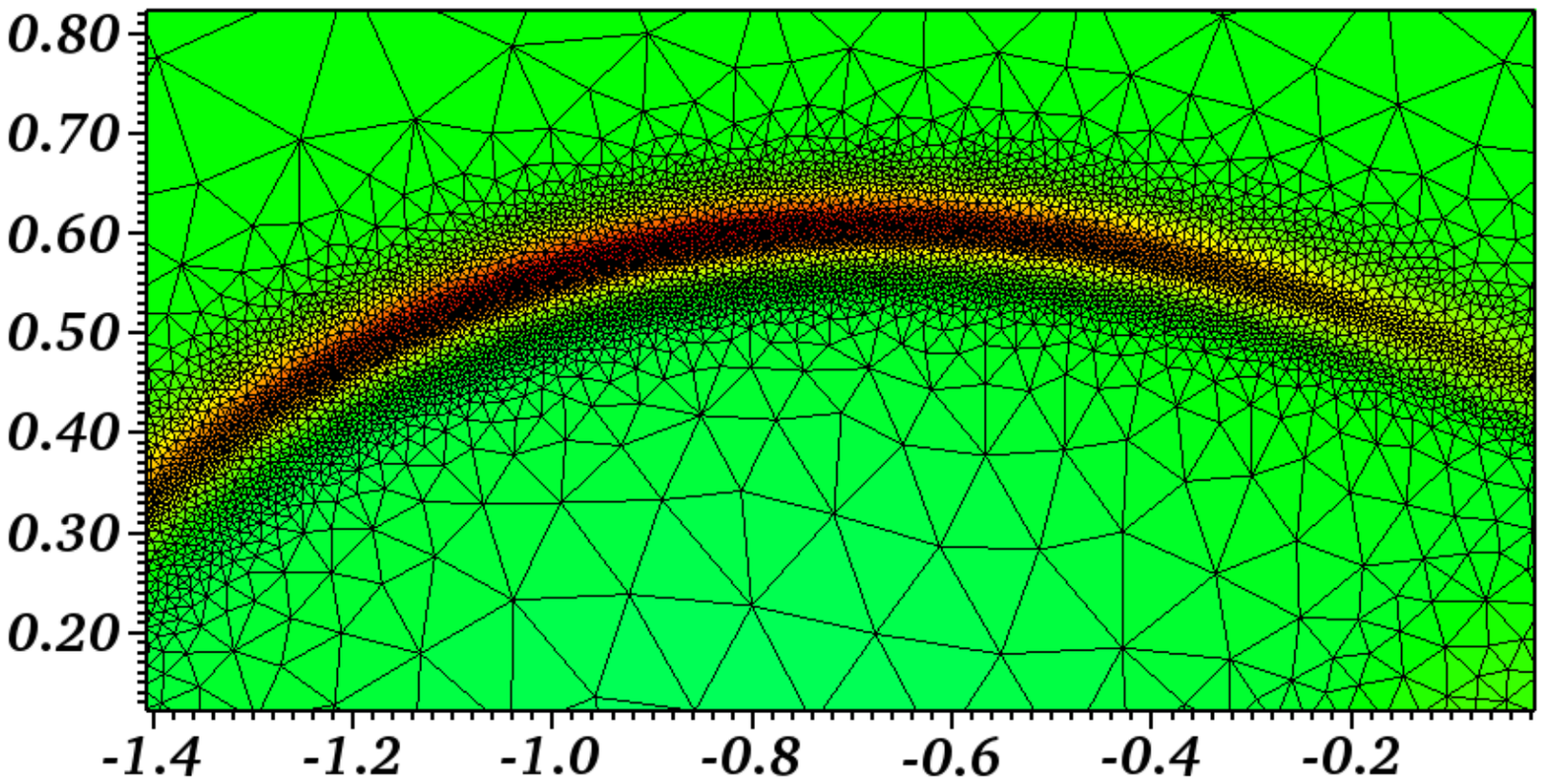}
  \caption{Zoom of current density at saturation ($t = 8.35$) of the tilt instability for $\eta = 0.0025$ (i.e.  $S  \simeq 1080$), overlaid with magnetic field lines (left panel), and overlaid 
  with the adapted mesh (right panel).
  }
\end{figure}
 
 \begin{figure}
\centering
 \includegraphics[scale=0.37]{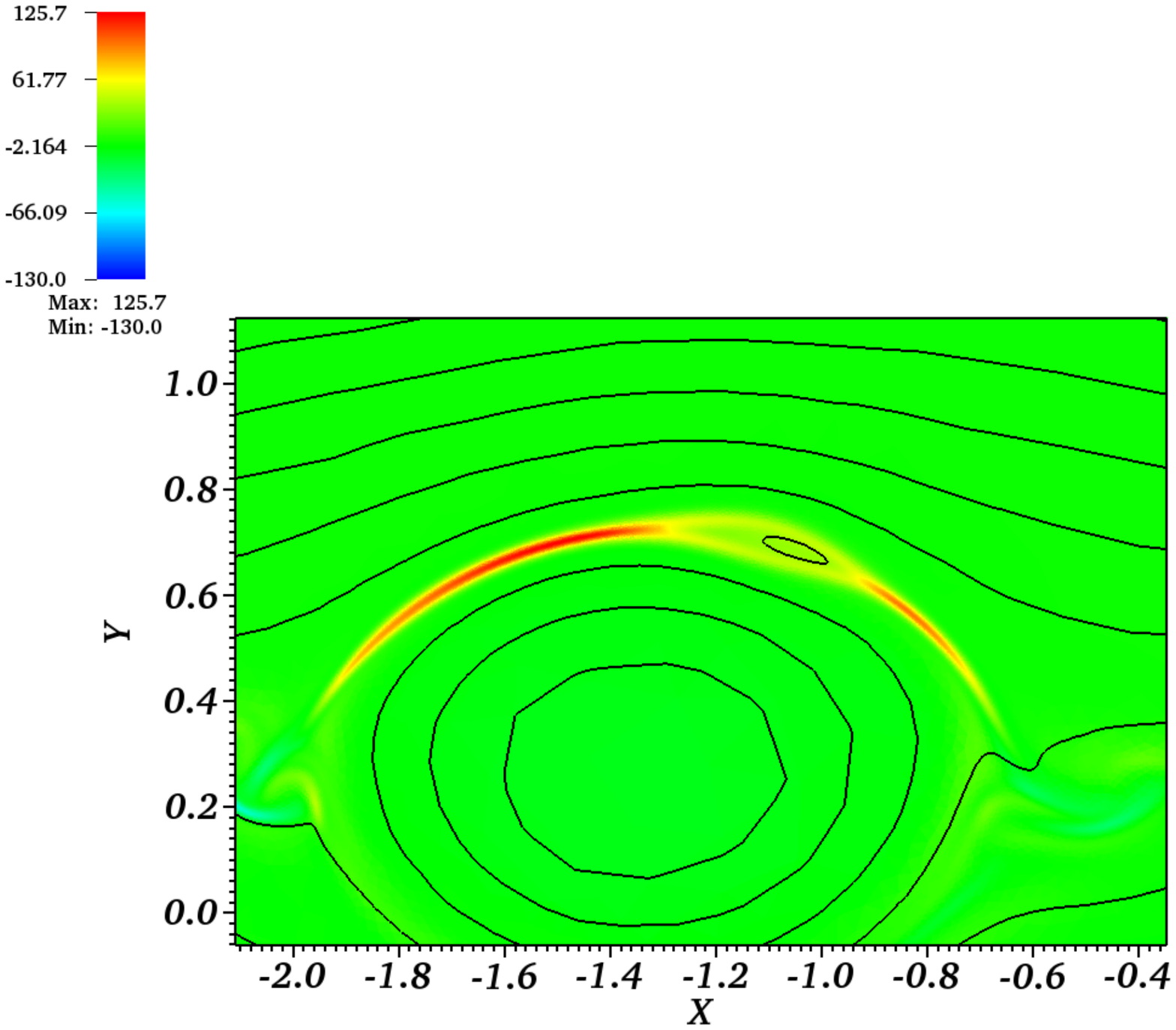}
 \includegraphics[scale=0.37]{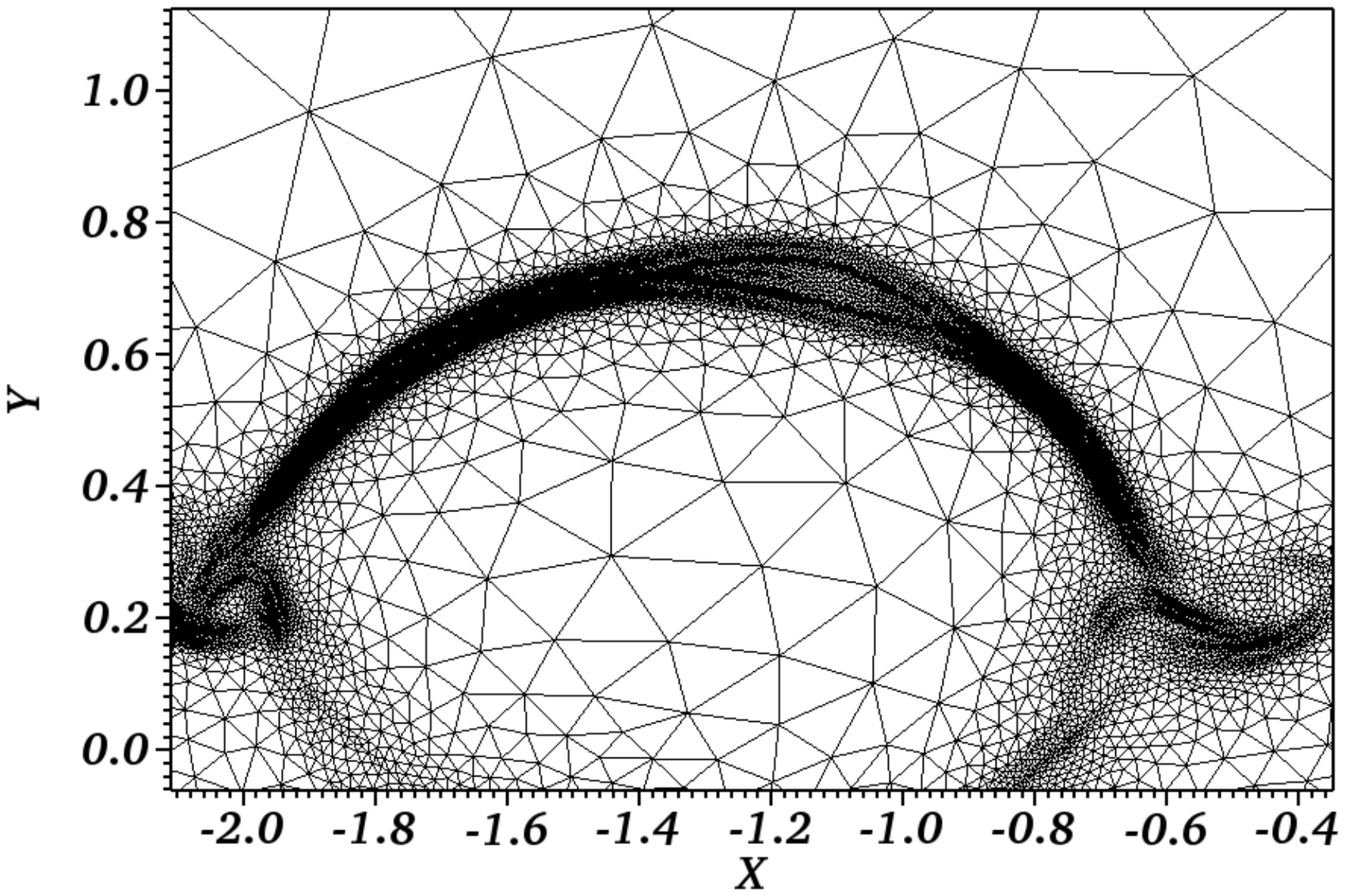}
  \caption{Zoom of current density (left current layer) for $S  \simeq 8000$ case at a relatively late time $t = 10.7$ during magnetic reconnection (see Figure 12)
  showing the transient formation of one plasmoid (left panel), and the corresponding adapted mesh (right panel).
    }
\end{figure}

 \subsubsection{Plasmoid reconnection regime}
We explore lower resistivity values in the range $[5 \times 10^{-4} : 1 \times 10^{-4}]$ , corresponding typically to $S$ values in the range $[8000: 40000]$.
The time history of the maximum
current density is plotted in Figure 15 for $4$ different $S$ values. First, we note that the formation of plasmoids occurs sooner when $S$
is higher. For example, for $S \simeq 13300$, plasmoids are seen to form close to the first peak ($t \simeq 9$). However, they can appear much
more earlier for $S \simeq 40000$ (i.e. at $t \simeq 8.3$). In the latter case, the two current sheets are invaded by a number of plasmoids
varying between $4$ and $6$, in rough agreement with previous studies showing that the number of plasmoids is an increasing function
of $S$. This exact number fluctuates during magnetic reconnection process because, new plasmoids are constantly forming, moving, eventually coalescing
(giving monster plasmoids), and finally being ejected through the outflow boundaries. These fluctuations also reflect the time oscillations
of the maximum current density seen in Figure 15. At a given time, the system thus appears as an aligned layer structure of plasmoids
of different sizes, as illustrated in Figure 16. 

A characteristic time scale for the plasmoids growth $\tau_p$ can be deduced from the slopes in Figure 15 for the maximum amplitude
of the current density, obtained just before the quasi-steady state involving fluctuations around an average value. This phase also corresponds
to a time interval, beginning when the expected SP value is attained and finishing when the plasmoids regime is fully obtained.
We have checked that the beginning also corresponds to the early apparition of the plasmoids (letter $P$ in Figure 15).
This time scale clearly decreases with $S$. For example, for the cases with $S \simeq 13300$, and $S \simeq 20000$, it
is much longer compared to the characteristic time scale of the tilt instability growth
(see the slope at a time around $t \simeq 7.5$) that is given by $\tau \simeq 0.38$. 
This is not the case for $S \simeq 80000$, as the two time scales have now comparable values $\tau_p \simeq \tau = 0.38$.
This scaling law dependence of $\tau_p$ decreasing with $S$ for $S  \simgt S_c$, 
agrees with previous numerical studies and also with theoretical models
where the growth rate of plasmoids for $S \simgt S_c$ increases with $S$  \citep{2017ApJ...850..142C, 2014ApJ...780L..19P}.

However, another critical Lundquist value ($S_T$) is predicted by \citep{2017ApJ...850..142C} in correspondence to a regime for which $\tau_p \ll \tau$, 
in disagreement with what is expected from the other model \citep{2014ApJ...780L..19P} for which $\tau_p$ remains of order
of the characteristic Alfv\'en time (i.e. $\tau_p  \simeq 1$ in our units) in the high Lundquist number  limit. 
This important point about the existence of this transitional Lundquist number ($S_T$)
will be explored in a future study, as higher $S$ values are required. Indeed, a value of $S_T  \simeq 10^6- 10^7$ is expected
from \citet{2017ApJ...850..142C}.

Another point of conflict between the two cited models concerns the critical aspect ratio $l/a$ (where $a$ is the
width of the current sheet that is thinning due to the tilt mode evolution) at which the plasmoids
begin to disrupt the current layer. Indeed, we have obtained that, this ratio is very close to the SP value (i.e. $l/a = S^{1/2}$),
in disagreement with the smaller value $l/a = S^{1/3}$ predicted by \citep{2014ApJ...780L..19P}.
For example, the case $S \simeq 40000$ gives
very different corresponding critical aspect ratio values of $200$ and $34$ for the two models respectively (see Figure 15).

Finally, we have plotted in Figure17 the maximum current density obtained at saturation as a function of $S$ for all the
runs. The results clearly show a transition between two regimes at a critical Lundquist $S_c \simeq 10^4$. 
Indeed, the values for $S  \simlt 10^4$ perfectly follow a Sweet-Parker scaling as $1.3 \times S^{1/2}$, while
another scaling linearly $S$ is obtained for $S  \simgt 10^4$. 
As concerns the reconnection rate, a constant reconnection rate of $\eta J_{max}  \simeq 0.05$ (independently of $S$) is obtained
in the plasmoid regime.
This is somewhat surprising as the fully plasmoids-dominated regime is probably not yet obtained for our moderately low $S$ values.
The normalized value $\eta J_{max}/(V_A B_u$) (where $B_u$ is the local magnetic field amplitude measured upstream of
the current sheet), is evaluated to be close to $0.012$, in rather good agreement with the value expected from the formula $10^{-2} (1 + P_r)^{-1/2} \simeq 0.014$
\citep{2016JPlPh..82f5901C}. This value also agrees with previously reported values of order $0.01$ in MHD simulations using coalescence instability setup
\citep{2009PhRvL.103j5004S, 2009PhPl...16k2102B}.

\begin{figure}
\centering
 \includegraphics[scale=0.37]{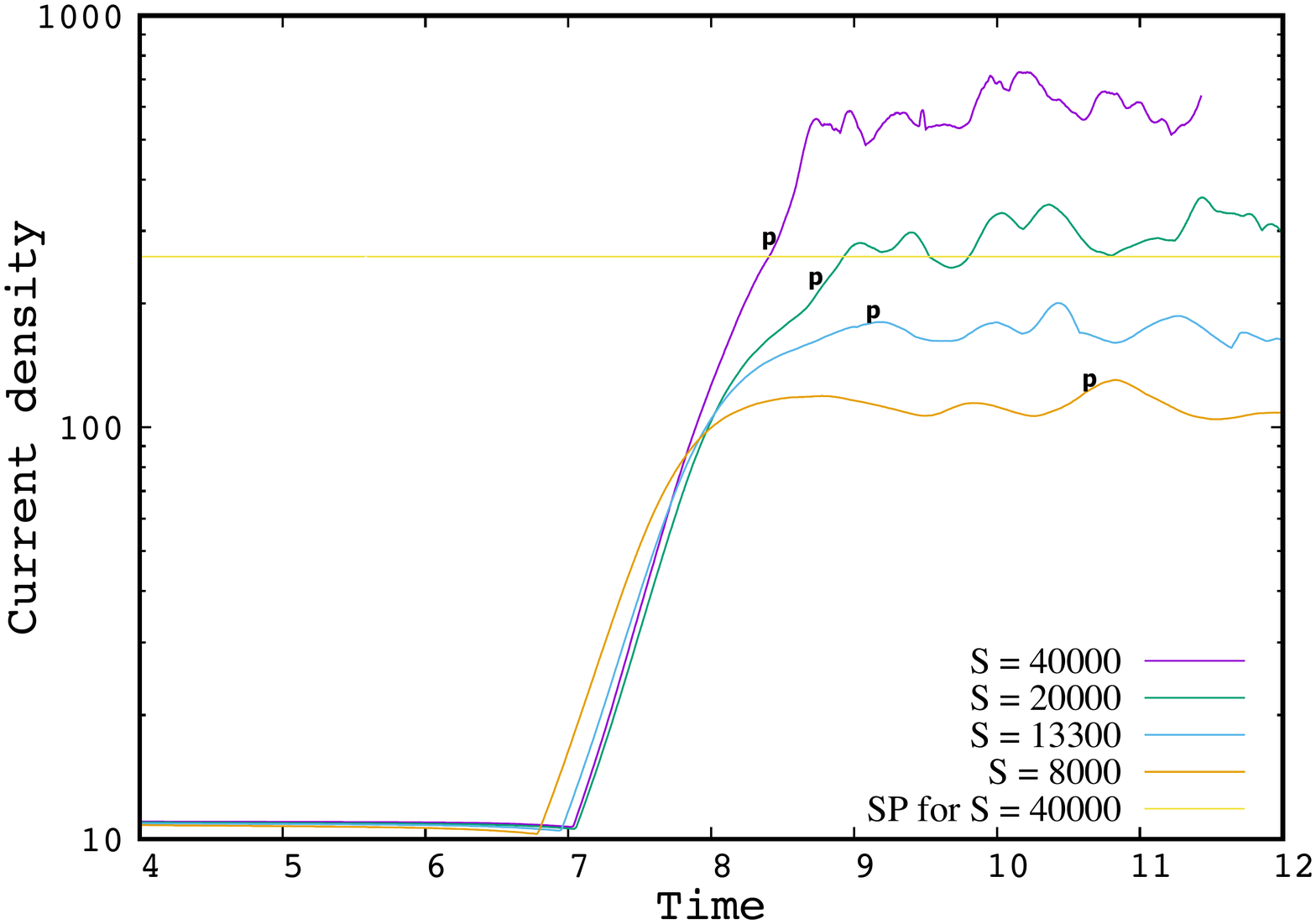}
  \caption{Time history of the maximum
  current density for four relatively high Lundquist (cases involving the formation of plasmoids). The Sweet-Parker value expected at saturation 
  for the maximum current density value (i.e. $S^{1/2}$)
  for the case $S  \simeq 40000$ is plotted for comparison.The letter $P$ indicates the time of disruption of the current sheet due to the plasmoid
  formation.
  }
\end{figure}

\begin{figure}
\centering
 \includegraphics[scale=0.47]{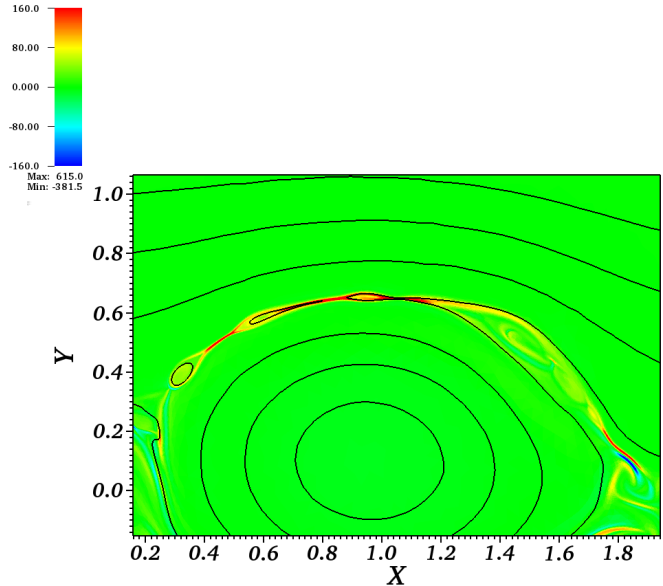}
  \caption{Zooms of current density at saturation of the tilt instability for $S  \simeq 40000$, overlaid with magnetic field lines. Note
  that saturated values in the range $[-160:160]$ are used in order to facilitate the visualisation of the current structures.
    }
\end{figure}

\begin{figure}
\centering
 \includegraphics[scale=0.47]{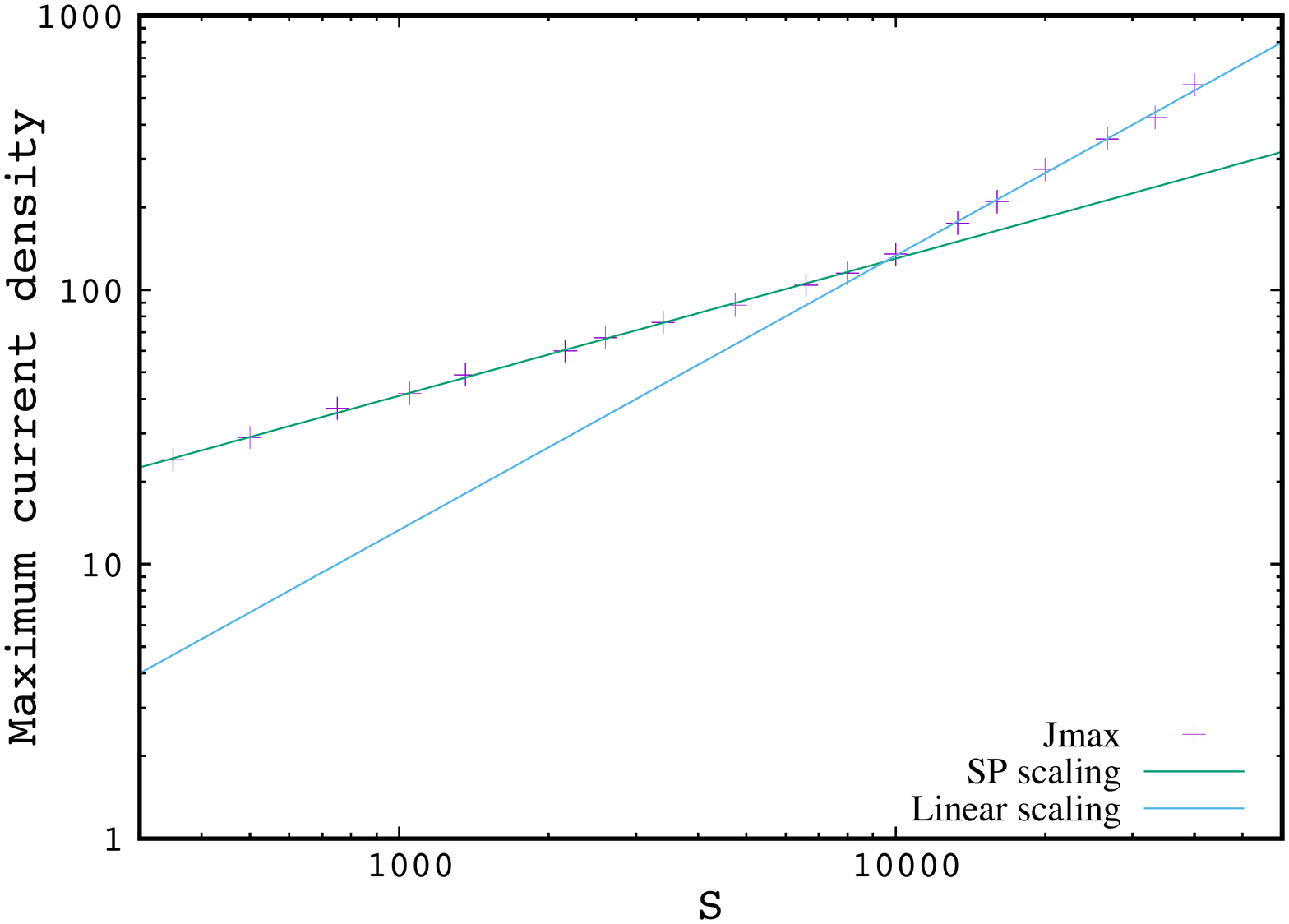}
  \caption{Maximum amplitude at saturation obtained for different Lundquist numbers. Sweet-Parker (SP) and linear scalings are also indicated for comparison.
  }
\end{figure}

\section{Summary, outlook, and astrophysical relevance}
\label{discussion_conclusion}

In this work, we have presented a new MHD code specifically designed to investigate magnetic reconnection
during the formation of quasi-singular current layers. In particular, the focus is on the regime of accelerated magnetic reconnection
that is characterized by disruption of the current sheets and the formation of chains of plasmoids. 
A 2D current-vorticity formulation is adopted, that is particularly suitable for our numerical approach.
A finite element method based on the use of quadratic finite element basis functions is taken, and is implemented
via a characteristic-Galerkin scheme. The complex time dependent structures are captured by the use of a highly adaptive
grid, based on the Hessian matrix of the solution (that is taken to be the density current for magnetic reconnection).

We can summarize our results as follows. 
First, we have applied our scheme to simplified test problems involving advection-diffusion and Burgers equations with small dissipation coefficients. 
Indeed, our approach has been demonstrated to be particularly efficient to reproduce the solutions of these
numerically challenging problems, where time-dependent quasi-singular structures can form.
Second, we have applied our scheme to the 2D set of MHD equations. The code, called FINMHD, has enabled us to follow
the development of the tilt instability in cartesian geometry. Contrary to the widely used coalescence instability, the tilt mode
produces in a more self-consistent way two quasi-singular current sheets in the domain. During the associated shrinking process,
an expected steady-state Sweet-Parker reconnection is driven if the Lundquist number $S$ of each layer is below a critical value $S_c \simeq 10^4$.
On the other hand, when $S \simgt S_c$, the reconnection is accelerated with a reconnection rate that is nearly independent of $S$.
In the latter case, the current layers are invaded by a number of plasmoids that is an increasing function of $S$. The maximum
amplitude of the density current exhibits somme oscillatory behavior with time with an average value
$J_{max}$ that scales as $S$ (contrary to the SP regime where it scales as $S^{1/2}$).
The results obtained for $10^4 \simlt S \simlt 10^5$ indicate that plasmoids can grow on a time scale $\tau_p$ that becomes of order of the characteristic time
of the current sheet formation $\tau$ ($\tau \simeq 0.38$ for the tilt instability) when $S$ is high enough. The SP aspect ratio of the current sheet
($l/a \simeq S^{1/2}$) is also approximately reached before the disruption time, in agreement with expectations from theoretical model of
\citet{2017ApJ...850..142C}.

Future studies employing higher $S$ values are clearly needed in order to investigate the existence of a second critical Lundquist number, $S_T$,
that is expected when the plasmoids growth time scale $\tau_p$ is much smaller than $\tau$. This is important in order to discriminate between
the two previously cited theoretical models. It is also important to use a wide range of different setups, like coalescence, tilt, and other instabilities, in order to
investigate the dependence with the shrinking process (including the time scale) of the forming current layers.
 
Such studies are of strong relevance in order to understand the mechanism at work in order to explain
the fast conversion of free magnetic energy in astrophysical environments such as in solar corona.
Given typical values of solar loop parameters in active regions, it is commonly admitted 
that the amount of available magnetic energy stored is sufficient to feed large flare events. However, 
observations require two fast time scales, that are given by the characteristic time scale of the whole process 
(with typical values of order of a few minutes) and an even faster time scale representative of an initial explosive-like phase. 
The reconnection rate of order $0.01 V_A B_u$, as obtained in the plasmoid dominated reconnection regime, is known
to give roughly the correct value for the first time scale. On the other hand, the second (explosive) time scale, that is
generally called the onset phase, is not clearly identified. The reconnection regime associated to the plasmoids growth,
where the reconnection rate increases abruptly on the explosive time scale $\tau_P$ (of order seconds) would thus
account for observations.

\
\ \

\textbf{Acknowledgments}\\
The author thank I. Moufid for fruitful work during its internship
that helped improve the choice of the numerical parameters in the FINMHD code.



\appendix

\section{Stability analysis of the schemes}
We consider the stability of a linearized version of Equation 15, where for the sake of simplicity the linearization is obtained
around a uniform equilibrium and the diffusions terms are dropped. In this way, the resulting linear system can be cast in the simple wave-like form,
\begin{equation}
    \left\{
    \begin{aligned}
    \frac{\partial \omega_1}{\partial t} =  (\bm{B_0} . \bm{\nabla})  J_1 , \\   
      \frac{\partial J_1}{\partial t} =  (\bm{B_0} . \bm{\nabla})   \omega_1 ,\\  
            \end{aligned}
    \right.
\end{equation}
where $\bm{B_0}$ is the equilibrium magnetic field vector, and $\omega_1$ and $J_1$ are the perturbed vorticity and current
density respectively. Indeed, the previous system is equivalent to,
\begin{equation}
    \left\{
    \begin{aligned}
    \frac{\partial ^2 J_1}{\partial t^2} =  V_A^2  \nabla^2 J_1 , \\   
     \frac{\partial ^2 \omega_1}{\partial t^2} =  V_A^2  \nabla^2\omega_1 ,\\ 
                \end{aligned}
    \right.
\end{equation}
where $V_A = B_0^2$ is the Alfv\'en speed.

The first order implicit time discretized form (following Equation 8) is,
\begin{equation}
    \left\{
    \begin{aligned}
     \frac{\omega_1^{n+1} - \omega_1^{n} } {\Delta t} =   (\bm{B_0} . \bm{\nabla})  J_1^{n+1} , \\  
     \dfrac{J_1^{n+1} - J_1^{n} } {\Delta t} =   (\bm{B_0} . \bm{\nabla})   \omega_1^{n+1} ,\\
            \end{aligned}
    \right.
\end{equation}
leading to the following amplification matrix,
\[
   T_1 =   \dfrac{1}  {1 + \alpha^2}
  \left[ {\begin{array}{cc}
   1 & i \alpha \\
   i \alpha & 1 \\
  \end{array} } \right] ,
\]
where $\alpha = k V_A \Delta t$ for a given wavenumber $k$. The associated (complex conjugate) eigenvalues are thus $(1 \pm i \alpha)/ (1 + \alpha^2)$ 
with a modulus equal to $1 / (1 + \alpha^2)^{1/2} $ always smaller than one. The scheme is consequently unconditionally stable,
and the spectrum of Alfv\'en waves are more or less damped depending on  $\alpha$. For this scheme, the choice of the time step is
only determined by the precision required on the linear growth rate of the tilt instability (that is of order unity). Typically, we use a time step $\Delta t = 10^{-2}$,
giving a very small second order damping factor of order $10^{-4}$ on one time step.

And the second order semi-implicit version (following Equation 9) is,
\begin{equation}
    \left\{
    \begin{aligned}
         \dfrac{\omega_1^{*} - \omega_1^{n} } {\Delta t/2} =   (\bm{B_0} . \bm{\nabla})  J_1^{*} \\  
     \dfrac{J_1^{*} - j_1^{n} } {\Delta t/2} =   (\bm{B_0} . \bm{\nabla})   \omega_1^{n} ,\\
       \end{aligned}
    \right.
\end{equation}
for the predictor step, and
\begin{equation}
    \left\{
    \begin{aligned}
        \dfrac{\omega_1^{n+1} - \omega_1^{n} } {\Delta t} =   (\bm{B_0} . \bm{\nabla})  J_1^{*} \\  
     \dfrac{J_1^{n+1} - j_1^{n} } {\Delta t} =   (\bm{B_0} . \bm{\nabla})   \omega_1^{*} ,\\
                 \end{aligned}
    \right.
\end{equation}    
for the corrector step. For this second scheme, the amplification matrix is,
\[
   T_2 =   
  \left[ {\begin{array}{cc}
   1 - \alpha ^2/2&  i \alpha \\
   i \alpha (1  - \alpha ^2/4 ) &   1 - \alpha ^2/2 \\
  \end{array} } \right] .
\]
The associated eigenvalues are thus $(1 -  \alpha ^2/2) \pm i \alpha (4 - \alpha^2)/2$ 
with a modulus equal to $1$. This second scheme is conditionally stable, as $ \alpha < 2$ is required.
This translates into a necessary CFL condition for stability, $\Delta t < 2/k_M$, where $k_M$ is the maximum numerical wavenumber
included in the simulation (typically $2 \pi /h_{min}$, with $h_{min}$ being the smallest triangle length scale).
For the simulations presented in this work, we have checked that stability is effectively obtained when
$\Delta t \simlt 10^{-3}$.

\section{Convergence of FINMHD with time step}
The system of reduced MHD equations (Equations 5-8) are integrated using our first-order (in time) characteristic-Galerkin
scheme (Equation 15) for the tilt instability problem using $\eta = \nu = 0.0025$. Note that $P_2$ elements are necessary in this case, as
the maximum spatial derivative in the MHD equations is second order. In order to save some CPU time computation, the adaptation
mesh procedure is also done only every $5$ time steps.
This scheme is checked to be linearly
inconditionally stable (see Appendix A). We have employed different time steps in different runs in order to study the convergence
of the code. The results are shown in Figure 16 for the maximum amplitude of the current density ($J_{max}$) and vorticity as functions
of time, for $5$ different initial time steps $\Delta t_{0}$. The convergence is shown to be obtained for $\Delta t_{0} \simeq 0.01$. 
We recall that in all runs, the time step is not constant in time, as it is adapted following $\Delta t = \Delta t_{0} J_e/J_{max}$, where
$J_e$ and $J_{max}$ are the maximum current density at equilibrium and current time respectively.
This is also confirmed in Figure 17, where is plotted the first peak amplitude of $J_{max}$ as a function of the inverse initial time step $1/\Delta t_{0}$.

A similar (slightly faster) convergence can be obtained by using the second-order scheme (Equations 16-17), that however is checked to be only
conditionally stable in agreement with the CFL criterion previously derived (see Appendix A). For the simulations presented in this work, we prefer
to use the first-order scheme, as we found the stability limitation to be too restrictive (a initial minimum length scale of $10^{-2}$ leading to
a maximum initial time step of order $10^{-3}$).

 \begin{figure}
\centering
 \includegraphics[scale=0.29]{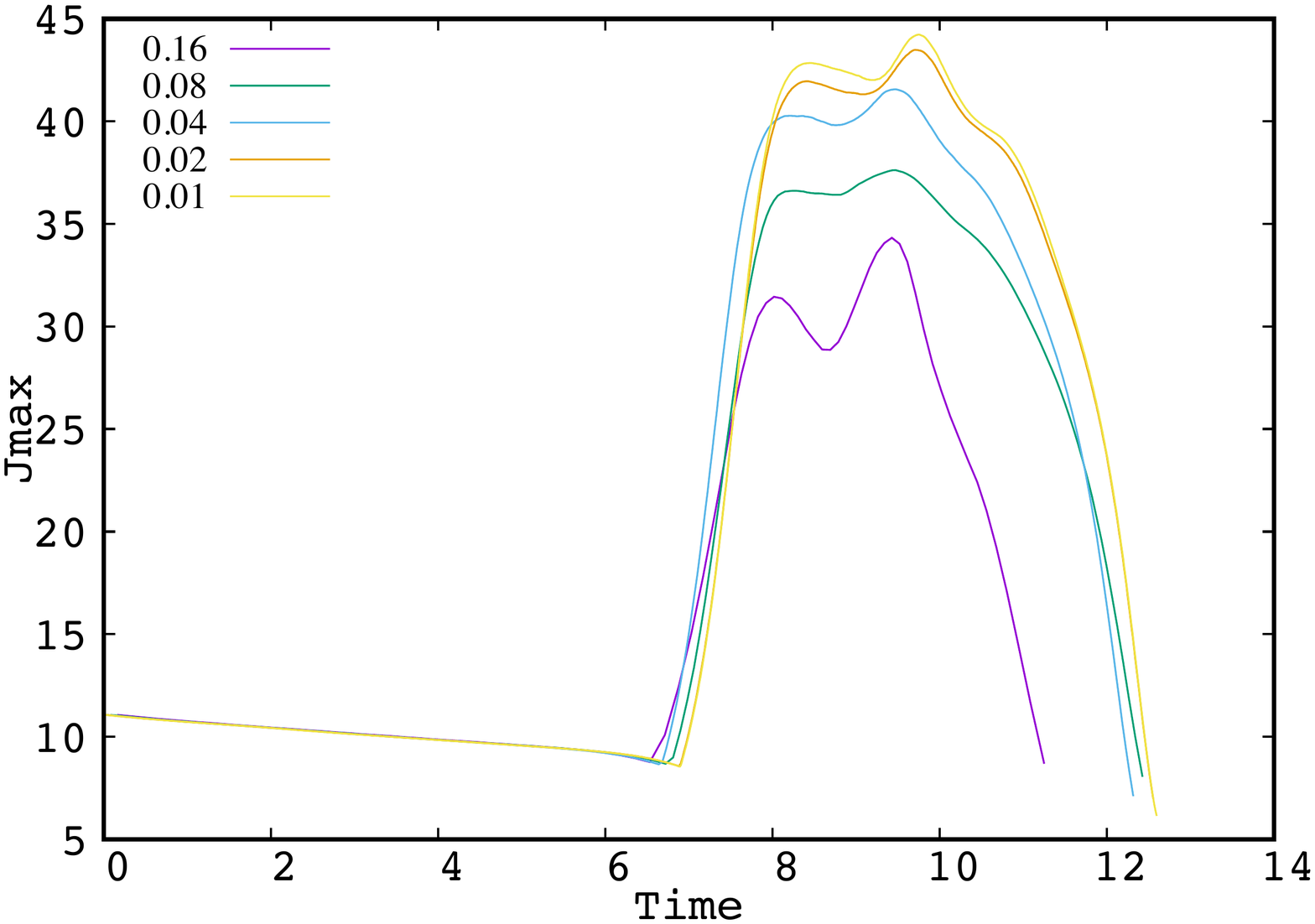}
 \includegraphics[scale=0.29]{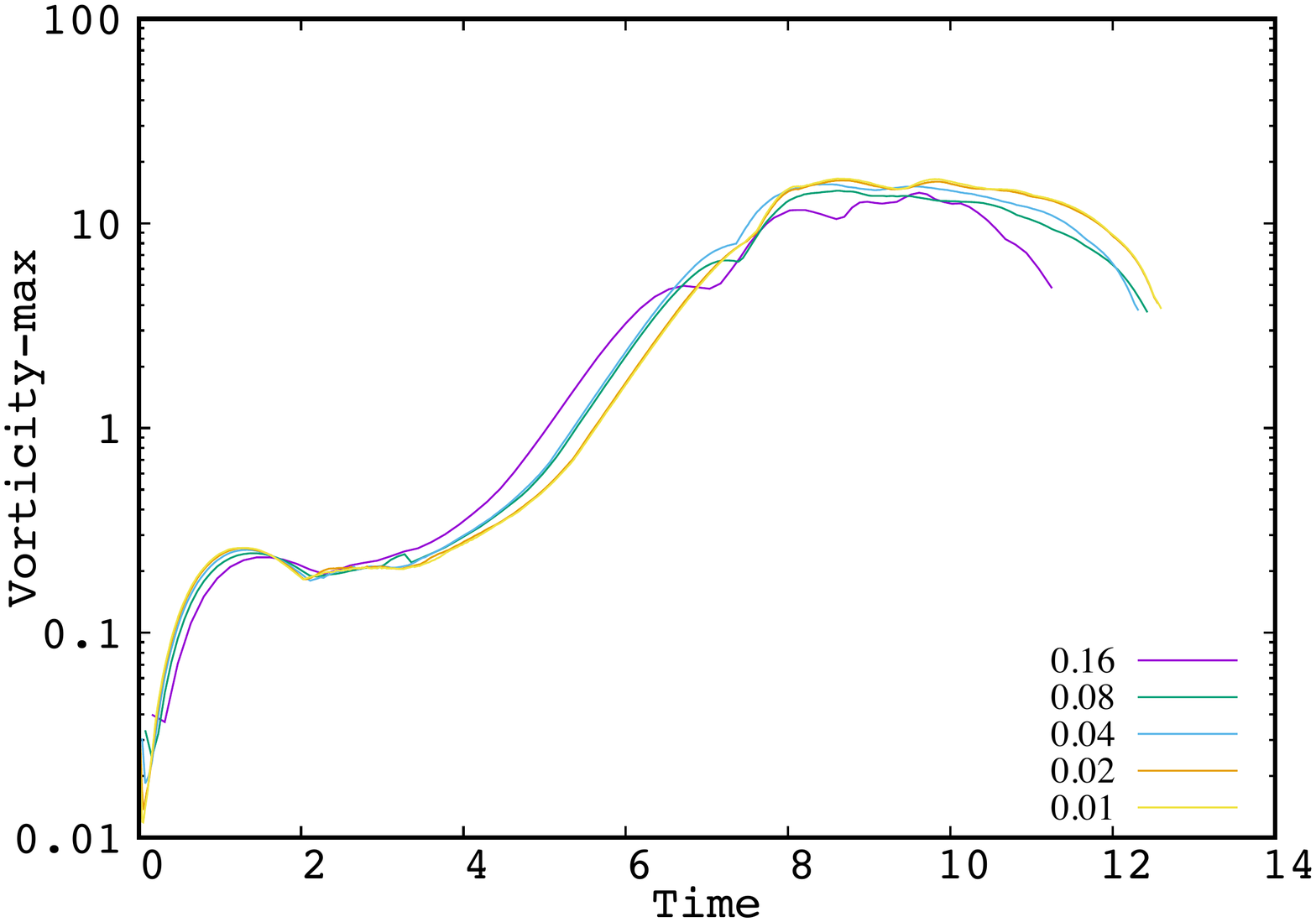}
  \caption{Maximum current density amplitude $J_{max}$ (left panel), and maximum vorticity amplitude (right panel) and as a function of time
  for $5$ runs using different values of the initial time-step $\Delta t_{0}$.
 }
\end{figure}

\begin{figure}
\centering
 \includegraphics[scale=0.5]{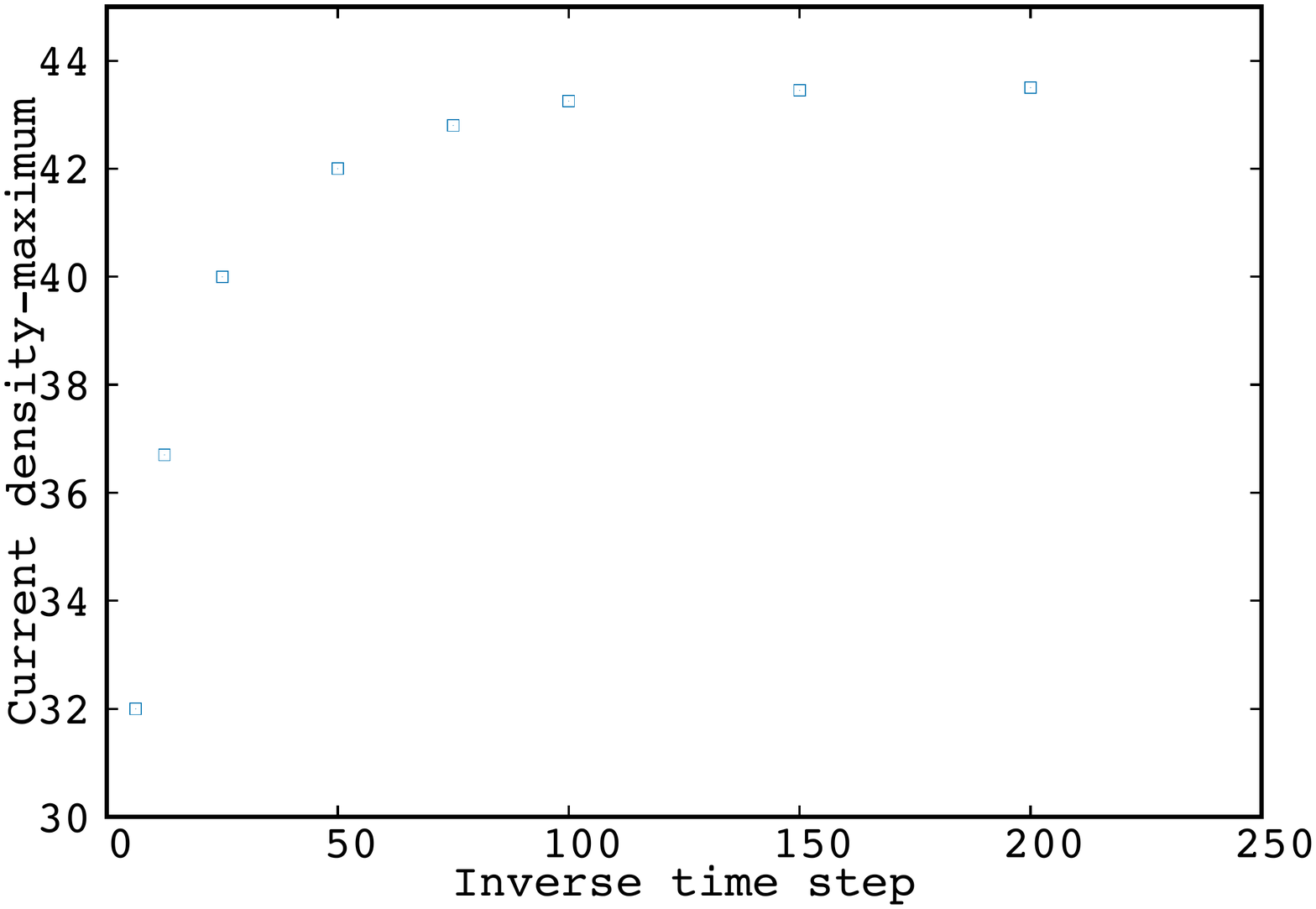}
  \caption{Maximum current density amplitude measured at the first peak (see previous figure) obtained for different runs
  as a function of the inverse initial time-step $1/\Delta t_{0}$.
   }
  \end{figure}


\begin{thebibliography}{}
\bibitem[Baty \& Nishikawa(2016)]{2016MNRAS.459..624B} Baty, H., \& Nishikawa, H.\ 2016, \mnras, 459, 624
\bibitem[Baty(2017)]{2017ApJ...837...74B} Baty, H.\ 2017, \apj, 837, 74
\bibitem[Bhattacharjee et al.(2009)]{2009PhPl...16k2102B} Bhattacharjee, A., Huang, Y.-M., Yang, H., \& Rogers, B.\ 2009, Phys. Plasmas, 16, 112102
\bibitem[Chen et al.(2013)]{che13} Chen, W, Ching-Shan, C., \& Chiu-Yen, K.\ 2013, Journal of Computational Physics, 234, 452
\bibitem[Comisso \& Bhattacharjee(2016)]{2016JPlPh..82f5901C} Comisso, L., \& Bhattacharjee, A..\ 2016, J. Plasma Phys., 13, 032307
\bibitem[Comisso et al.(2017)]{2017ApJ...850..142C} Comisso, L., Lingam, M., Huang, Y.~M., \& Bhattacharjee, A.\ 2017, \apj, 850, 142
\bibitem[Hecht(2012)]{hec12} Hecht, F.\  2012, Journal of numerical mathematics 20, 251
\bibitem[Huang et al.(2017)]{2017ApJ...849...75H} Huang, Y.~M., Comisso, L., \& Bhattacharjee, A.\ 2009, \apj, 849, 75
\bibitem[Keppens et al.(2013)]{2013PhPl...20i2109K} Keppens, R., Porth, O., Galsgaard, K., et al.\ 2013, Phys. Plasmas, 20, 092109
\bibitem[Keppens et al.(2014)]{2014ApJ...795...77K}  Keppens, R., Porth, O., \& Xia, C.\ 2014, \apj, 795, 77
\bibitem[Knoll \& Chac\'on(2006)]{2006PhPl...13c2307K} Knoll, D.~A., \& Chac\'on, L..\ 2006, Phys. Plasmas, 13, 032307
\bibitem[Lankalapalli et al.(2007)]{2007JCoPh.225..363L} Lankalapalli, S., Flaherty, J.~E., Shephard, M.~S., \& Strauss, H.~R.\ 2007, \jcp, 225, 363
\bibitem[Loureiro et al.(2007)]{2007PhPl...14j0703L} Loureiro, N.~F., Schekochihin, A.~A., \& Cowley, S.~C.\ 2007, Phys. Plasmas, 14, 100703
\bibitem[Manzinali et al.(2018)]{2018CMAME.340..864M} Manzinali, G, Hachem, E., \& Mesri, Y.\ 2018, Computer Methods in Applied Mechanics and Engineering, 340, 864
\bibitem[Ng et al.(2007)]{2008ApJS..177..613N} Ng, C.~W., Rosenberg, D., Germaschewski, K., Pouquet, A., \& Bhattacharjee, A.\ 2007, \apjs, 177, 613
\bibitem[Parker(1957)]{1957JGR...62...509} Parker, E.~N.\ 1957, \jgr, 62, 509
\bibitem[Philip et al.(2007)]{phi07} Philip, B., Pernice, M., \& Chacon, L.\ 2007, Lecture Notes in Computational Science and Engineering
55, 723-729
\bibitem[Pironneau(1992)]{1992CMAME.100..117P} Pironneau, O.\ 1992, Computer Methods in Applied Mechanics and Engineering, 100, 117
\bibitem[Porth et al.(2014)]{2014ApJS..214....4P} Porth, O., Xia, C., Hendrix, T., Moschou, S.~P., \& Keppens, R.\ 2014, \apjs, 214, 4
\bibitem[Priest \& Forbes(2000)]{pri00} Priest, E.~R., Forbes, T.~G.\ 2000, Magnetic Reconnection (Cambridge: Cambridge Univ. Press) 
\bibitem[Pucci \& Velli(2014)]{2014ApJ...780L..19P} Pucci, F., \& Velli, M.\ 2014, \apjl, 780, L19
\bibitem[Richard et al.(1990)]{1990PhFlB...2..488R} Richard, R.~L., Sydora, R.~D., \& Ashour-Abdalla, M.\ 1990, Phys. Fluids B, 225, 363
\bibitem[Ripperda et al.(2017)]{2017MNRAS.467.3279R} Ripperda, B., Porth, O., Xia, C., \& Keppens, R.\ 2017, \mnras, 467, 3279
\bibitem[Rui \& Tabata(2002)]{rui02} Rui, H. \& Tabata, M.\ 2002, Numer. Math., 92, 161
\bibitem[Samtaney et al.(2009)]{2009PhRvL.103j5004S} Samtaney, R., Loureiro, N.~F., Uzdensky, D.~A., Schekochihin, A.~A., \& Cowley, S.~C.\ 2009, 
Phys. Rev. Lett., 103, 105004l 
\bibitem[Strauss \& Longcope(1998)]{1998JCoPh.147..318S} Strauss, H.~R., \& Longcope, D.~W.\ 1998, \jcp, 147, 318
\bibitem[Sweet(1958)]{swe58} Sweet, P.~A.\ 1958, Electromagnetic Phenomena in Cosmical Physics (IAU Symposium vol 6, ed Lehnert B p 123)

\end{thebibliography}
\end{document}